\documentclass[11pt,times,letter]{article}
\usepackage{amsmath, amsfonts,amsthm}
\usepackage{dsfont}
\usepackage{bbm}

\usepackage[english]{babel}
\usepackage[normalem]{ulem}

\usepackage{latexsym,amssymb,graphicx}
\usepackage{float}
\usepackage{verbatim}
\usepackage{mathrsfs}
\usepackage{epsfig}

\usepackage{xcolor}
\usepackage{tikz}
\usepackage{pgfplots}
\usepgfplotslibrary{fillbetween}

\usepackage{url}
\usepackage{bm}
\usepackage[numbers]{natbib}

\usepackage[colorlinks,linkcolor=blue,citecolor=blue,anchorcolor=blue]{hyperref}

\newif\ifshowrevisions
\showrevisionstrue
\newif\ifkeeprevisioncolors
\newcommand*{\revisionink}[1]{\ifkeeprevisioncolors\color{#1}\else\color{black}\fi}

\definecolor{Orange}{RGB}{226,118,0}

\definecolor{RevTeal}{RGB}{0,135,155}

\definecolor{RevMagenta}{RGB}{214,0,128}

\definecolor{RevJointBlue}{RGB}{0,70,160}

\newenvironment{jointblock}{\ifshowrevisions\revisionink{RevJointBlue}\fi}{}
\definecolor{RevBassViolet}{RGB}{96,48,160}

\definecolor{RevBorelBrown}{RGB}{125,72,32}
\DeclareRobustCommand{\borelrev}[1]{\ifshowrevisions{\revisionink{RevBorelBrown}#1}\else{#1}\fi}
\newenvironment{borelblock}{\ifshowrevisions\revisionink{RevBorelBrown}\fi}{}
\definecolor{RevSMOlive}{RGB}{104,114,0}

\definecolor{RevMultiTurquoise}{RGB}{0,105,85}
\DeclareRobustCommand{\drev}[1]{\ifshowrevisions{\revisionink{RevMultiTurquoise}#1}\else{#1}\fi}

\definecolor{RevConvergenceWine}{RGB}{145,30,65}
\DeclareRobustCommand{\crev}[1]{\ifshowrevisions{\revisionink{RevConvergenceWine}#1}\else{#1}\fi}
\newenvironment{convergenceblock}{\ifshowrevisions\revisionink{RevConvergenceWine}\fi}{}
\definecolor{RevBasketIndigo}{RGB}{60,75,180}

\newenvironment{basketblock}{\ifshowrevisions\revisionink{RevBasketIndigo}\fi}{}
\definecolor{RevisionPurple}{RGB}{110,55,135}

\newtheorem{theorem}{Theorem}[section]
\newtheorem{corollary}[theorem]{Corollary}
\newtheorem{lemma}[theorem]{Lemma}
\newtheorem{proposition}[theorem]{Proposition}

\newtheorem{remark}[theorem]{Remark}
\newtheorem{example}[theorem]{Example}

\numberwithin{equation}{section}

\definecolor{Red}{rgb}{1.00, 0.00, 0.00}

\definecolor{DRed}{rgb}{0.5, 0.00, 0.00}

\definecolor{Blue}{rgb}{0.00, 0.00, 1.00}

\definecolor{Green}{rgb}{0.0, 0.4, 0.0}

\definecolor{Gray}{rgb}{0.7421875,0.7421875,0.7421875}

\definecolor{mycolor1}{rgb}{0.00000,0.44700,0.74100}%
\definecolor{mycolor2}{rgb}{0.85000,0.32500,0.09800}%
\definecolor{mycolor3}{rgb}{0.92900,0.69400,0.12500}%
\definecolor{mycolor4}{rgb}{0.49400,0.18400,0.55600}%
\definecolor{mycolor5}{rgb}{0.46600,0.67400,0.18800}%
\definecolor{mycolor6}{rgb}{0.00000,1.00000,1.00000}%

\allowdisplaybreaks[4]
\pgfplotsset{compat=newest}

\def\beq{\begin{eqnarray}}
\def\eeq{\end{eqnarray}}
\def\be*{\begin{eqnarray*}}
\def\ee*{\end{eqnarray*}}

\def \E{\mathbb{E}}

\def \I{\mathbb{I}}

\def \Q{\mathbb{Q}}
\def \R{\mathbb{R}}

\def\Fc{{\cal F}}

\def\Pc{{\cal P}}

\def\Wc{{\cal W}}

\def \cv {\le_{\rm c}}

\def\E{\mathbb{E}}

\def\Pc{\mathcal{P}}

\def\R{\mathbb{R}}

\def\Fc{\mathcal{F}}

\def\Hh{\mathcal{H}}

\def\namedlabel#1#2{\begingroup
	#2%
	\def\@currentlabel{#2}%
	\phantomsection\label{#1}\endgroup
}

\def \1 {{\mathbf 1}}

\title{\Large \bf
Price-impact Martingale optimal transport
}
\date{\vspace{-5ex}}

\author{%
  Erhan Bayraktar\thanks{{\small University of Michigan,
    Department of Mathematics (\mbox{erhan@umich.edu}).}}%
  \qquad \
  Shuoqing Deng\thanks{{\small The Hong Kong University of Science and Technology,
    Department of Mathematics (\mbox{masdeng@ust.hk}).}}%
  \qquad \
  Daxin Huang\thanks{{\small The Hong Kong University of Science and Technology,
    Department of Mathematics (\mbox{dhuangax@connect.ust.hk}).}}%
  \qquad  \ 
  Dominykas Norgilas\thanks{{\small North Carolina State University,
    Department of Mathematics (\mbox{dnorgil@ncsu.edu}).}}%
}

\begin{document}


\maketitle

\begin{abstract}

Martingale optimal transport (MOT) provides model-independent bounds for option prices calibrated to vanilla options, assuming frictionless trading in the underlying. In this work, we consider the case where dynamic trading incurs quadratic temporary price impact. First, we prove a duality result for the superhedging price. In particular, the primal of the duality relaxes the strict martingale constraint by penalizing the deviation from a specific martingale. Second, for one asset and one trading period, we interpret this price-impact martingale optimal transport (PIMOT) as a weak optimal transport (WOT) problem, and provide a characterization in the spirit of Wasserstein projection. Finally, we justify that PIMOT is an interpolation between MOT and classical optimal transport (OT), and derive the convergence rate to both sides. Numerical examples with SPX option data show how the price bounds per claim depend on the order size.

\vspace{0.6 cm}


\vspace{0.2 cm}



\end{abstract}

\section{Introduction}

Martingale optimal transport (MOT) was introduced in discrete time by Beiglb\"ock, Henry-Labord\`ere and Penkner~\cite{BHP13} and in continuous time by Galichon, Henry-Labord\`ere and Touzi~\cite{GHLT14} to compute model-independent bounds for option prices. As Breeden and Litzenberger~\cite{BL78} observed, call prices at all strikes determine the law of the underlying at their maturity. With two maturities, a model that is consistent with the quotes and free of arbitrage is therefore a coupling of the laws $\mu_0$ and $\mu_1$ of the prices $X_0$ and $X_1$ under which the price is a martingale. The upper bound for a claim $\Phi(X_0,X_1)$ is
\[
 \sup_{\Q\in\mathcal M(\mu_0,\mu_1)}\E_\Q[\Phi(X_0,X_1)],
 \qquad
 \mathcal M(\mu_0,\mu_1):=\bigl\{\Q\in\Pi(\mu_0,\mu_1):\E_\Q[X_1\mid X_0]=X_0\bigr\},
\]
where $\Pi(\mu_0,\mu_1)$ is the set of couplings of $\mu_0$ and $\mu_1$. For integrable marginals, Strassen's theorem~\cite{Strassen1965} shows that $\mathcal M(\mu_0,\mu_1)$ is nonempty exactly when $\mu_1$ dominates $\mu_0$ in convex order. 
The dual problem asks for the cheapest semi-static superhedge, which combines vanilla options traded at time zero with dynamic trading in the underlying, and~\cite{BHP13} show that the two values coincide under suitable conditions.

Classical MOT treats both parts of the hedge as frictionless. Liang et al.~\cite{BAMOT26} introduce bid--ask MOT, which brings the bid--ask spreads of vanilla options into this framework. They show that the resulting value is the superhedging price when options are bought at the ask and sold at the bid, and they quantify its convergence to MOT as the spreads vanish. Frictions can also enter through the dynamic part of the hedge, as costs of trading the underlying. Dolinsky and Soner~\cite{DS14} charge proportional costs to both the options and the stock, and their dual problem imposes approximate marginal and martingale conditions. Cheridito, Kupper and Tangpi~\cite{CKT17} allow convex costs for dynamic trading and keep the marginal constraints of MOT when call bid and ask prices coincide.

The two parts of a semi-static hedge are executed differently. Consider a dealer who sells a large exotic claim on an underlying with low trading volume. The static hedge is a single trade at time zero, of known size and at a price agreed per contract, so the dealer takes the option prices as given. The dynamic hedge is rebalanced at every fixing date in path-dependent amounts that must be executed promptly, and temporary impact is the cost of this immediacy. Gu\'eant and Pu~\cite{GueantPu17} point out that such costs cannot be ignored when the nominal of the option is large compared with the volume traded in the underlying. Their model and that of Bank, Soner and Vo{\ss}~\cite{BSV17} charge these costs to the trades in the underlying. A spread on the options is paid once and grows linearly in the quantity, while temporary impact grows quadratically in the size of each order and recurs at every rebalancing. We therefore keep the quoted call prices for the static part and charge impact to the dynamic part only.

In this paper, we consider semi-static superhedging with $d$ assets and fixing dates $0,\ldots,N$. Call prices determine the marginal law $\mu_i^a$ of asset $a$ at date $i$, so a static payoff $\varphi_i^a(X_i^a)$ built from calls and forwards costs $\int\varphi_i^a\,d\mu_i^a$. The dynamic part holds $H_i^a$ shares of asset $a$ from date $i$ to date $i+1$, starting and ending with no position, and an order of $\Delta H_i^a:=H_i^a-H_{i-1}^a$ shares pays the impact cost $\kappa_i^a(\Delta H_i^a)^2/2$. The superhedging price of a claim $\Phi$ is
\[
 D_\kappa(\Phi):=\inf\biggl\{\sum_{i,a}\int\varphi_i^a\,d\mu_i^a:\
 \sum_{i,a}\varphi_i^a(X_i^a)+\sum_{i<N,\,a}H_i^a\,(X_{i+1}^a-X_i^a)-\sum_{i,a}\frac{\kappa_i^a}{2}(\Delta H_i^a)^2\geq\Phi\biggr\},
\]
where the inequality holds on every price path. We study this lowest initial cost of a superhedge and call the pricing problem price-impact MOT (PIMOT).

Theorem~\ref{thm:main}, our first main result, gives PIMOT a transport formulation for upper semicontinuous claims of linear growth, with any number of assets and fixing dates. It shows that $D_\kappa(\Phi)=\sup_{\Q}\{\E_\Q[\Phi]-I_\kappa(\Q)\}$, where $\Q$ ranges over the laws with marginals $\mu_i^a$ and the penalty $I_\kappa(\Q)$ is the largest expected trading gain under $\Q$, net of impact costs. Without friction, this gain is zero under martingale laws and infinite otherwise, which recovers the duality of MOT. With quadratic impact, it is finite for laws with square-integrable conditional drifts and measures, as a weighted quadratic distance, how far the price process is from a specific martingale. As in frictionless MOT, there is no semi-static arbitrage if and only if the marginals of each asset increase in convex order (Proposition~\ref{prop:convex-order-na}).

Our second contribution concerns the one-asset, one-period problem. Proposition 3.1 formulates PIMOT as weak optimal transport \cite{ABP21, BBP19, GRST17} with a quadratic penalty on the conditional drift. Theorem 3.2 shows that, whenever the value is finite, all optimal couplings share the same drift and the objective gap controls the squared \(L^2(\mu_0)\) distance to that drift. For square-integrable marginals, the theorem also gives an equivalent optimization over intermediate laws in the convex-order set used in Wasserstein projection \cite{ACJ19, GJ18}. The unique optimal intermediate law is the law of \(\mathbb E_\gamma[X_1\mid X_0]\) under any optimal coupling \(\gamma\).

Finally, when all impact coefficients are multiplied by $\varepsilon>0$, the PIMOT price decreases to the MOT value as $\varepsilon\downarrow0$ and increases to the OT value as $\varepsilon\to\infty$ (Theorem~\ref{thm:asy-endpoints}). The gaps are of order $1/\varepsilon$ at the OT end and, under a bounded frictionless superhedge, of order $\varepsilon$ at the MOT end (Propositions~\ref{prop:rate-OT} and~\ref{prop:rate-additive}). Since $n$ claims at impact $\kappa$ have the same price per unit as one claim at impact $n\kappa$, the price interval per claim widens from the MOT interval towards the OT interval as the order grows. Section~\ref{sec:fin} illustrates this with forward-start and basket calls, including SPX data.

\textbf{Organization of the paper.} Section~\ref{sec:main} introduces the market with quadratic temporary impact, proves the duality and characterizes the absence of arbitrage across maturities. Section~\ref{sec:one-period} treats one asset and one trading period, derives the weak transport formulation and describes the optimal couplings. Section~\ref{subsec:asymptotics} derives the OT and MOT limits and their convergence rates. Section~\ref{sec:fin} computes price bounds for the forward-start and basket calls. Section~\ref{sec:proof} contains the proofs, and Appendix~\ref{app:fin-calibration} describes the calibration and the numerical implementation.

\subsection{Notation}\label{subsec:notation}

Throughout this paper, $\langle x,y\rangle$ denotes the inner product of $x,y\in\R^d$, $|x|$ the Euclidean norm and $\|x\|_1:=\sum_{a=1}^d|x^a|$. We abbreviate histories as $x_{0:i}:=(x_0,\ldots,x_i)$, for points as well as for random vectors. The positive and negative parts of $r\in\R$ are $r^+:=\max\{r,0\}$ and $r^-:=\max\{-r,0\}$.

Let $A$ be a closed subset of a Euclidean space. We write $\mathcal P(A)$ for the set of Borel probability measures on $A$, equipped with the weak topology induced by the bounded continuous functions on $A$, $\delta_x$ for the Dirac measure at $x\in A$, and, for $p\geq1$,
$$\Pc_p(A):=\left\{\nu\in\Pc(A):\int_A|x|^p\,\nu(dx)<\infty\right\}.$$
A Borel map $f$ on $A$ sends $\nu\in\mathcal P(A)$ to its image measure $f_\#\nu:=\nu\circ f^{-1}$. Couplings of $\nu\in\mathcal P(A)$ and $\nu'\in\mathcal P(A')$ are the probability measures on $A\times A'$ with marginals $\nu$ and $\nu'$, and $\Pi(\nu,\nu')$ denotes their set. On $\mathcal P_2(A)$, the quadratic Wasserstein distance is
$$\Wc_2(\nu,\nu')^2:=\inf_{\pi\in\Pi(\nu,\nu')}\int_{A\times A}|x-y|^2\,\pi(dx,dy).$$
Two laws $\nu,\nu'\in\mathcal P_1(\R)$ satisfy $\nu\cv\nu'$, and $\nu'$ is said to dominate $\nu$ in convex order, if $\int\psi\,d\nu\leq\int\psi\,d\nu'$ for every convex function $\psi:\R\to\R$.

The space $L^p(\nu;\R^d)$ consists of the $\R^d$-valued Borel functions $f$ with $\int|f|^p\,d\nu<\infty$, and $L^p(\nu):=L^p(\nu;\R)$. The same notation is used for a law $\Q$ on path space, with $\E_\Q$ the expectation under $\Q$. If $F$ is a closed subset of a Euclidean space, $C_b(F)$ is the space of bounded continuous functions on $F$, and $C_1(F)$ consists of the continuous functions $f$ on $F$ with $|f(x)|\leq C(1+|x|)$ for some $C<\infty$.

\section{Model and Duality} \label{sec:main}

We consider a semi-static market in which vanilla options can be traded at exogenous quotes, while dynamic trading in the underlying incurs temporary price impact. All prices are expressed in units of cash, with zero interest and no dividends. The static part of the hedging strategy is set up once, while the dynamic part is rebalanced at every fixed date in amounts that depend on the realized path. As motivated in the Introduction, we charge execution costs to the dynamic part only. Section~\ref{subsec:market} describes the market, Section~\ref{subsec:superhedging} formulates the superhedging problem and its primal counterpart, and Section~\ref{subsec:duality} states the duality; Section~\ref{subsec:calibration-na} treats no-arbitrage.

\subsection{The market}\label{subsec:market}

As in \cite[Section~2]{EGLO20}, calls are quoted on individual assets, and a position in any one asset may depend on the observed history of the entire price vector. In our market, the temporary impact on each asset's execution price depends only on the size of the order in that asset.

Fix $N,d\geq1$ and set $E:=\R^d$ and $\Omega:=E^{N+1}$. At each fixed date $i=0,\ldots,N$, we write $X_i=(X_i^1,\ldots,X_i^d)$ for the prices of the $d$ assets, and $\Fc_i:=\sigma(X_0,\ldots,X_i)$
is the information available at that date. The canonical process $X=(X_i)_{i=0}^N$ records the unaffected reference prices and provides the fixings for cash-settled contracts.

The canonical process $X=(X_i)_{i=0}^N$ records the unaffected reference prices used to calculate the payoffs of cash-settled contracts.

The static instruments are cash, calls and cash-settled forwards. All static positions are established initially, and the quoted unit prices apply to purchases and sales of any size. For each asset $a=1,\ldots,d$ and maturity $i=1,\ldots,N$, calls are quoted at every strike $K\in\R$. The call pays $(X_i^a-K)^+$ at date $i$. We assume that, for some $\mu_i^a\in\mathcal P_1(\R)$, its initial price is
$$C_i^a(K)=\int_{\R}(x-K)^+\,\mu_i^a(dx),\qquad K\in\R.$$
By the Breeden--Litzenberger relation \cite{BL78}, the call curve determines $\mu_i^a$ uniquely. in the sense of distributions. If $X_0$ is a future fixing, calls are also quoted for $i=0$, and static positions are chosen before $X_0$ is observed. For a known spot vector $x_0$, we use the same call-price formula at $i=0$ with $\mu_0^a=\delta_{x_0^a}$ for each asset $a$.

For each asset $a$ and fixing $i$, let $F_i^a:=\int_{\R}x\,\mu_i^a(dx)$ denote the mean implied by the call curve. At each quoted maturity $i$, we assume that a cash-settled forward on asset $a$ with delivery price $F_i^a$ is available at zero initial cost. A long position pays
$X_i^a-F_i^a$ at maturity. The forward provides static exposure to $X_i^a$, while buying the asset at date $0$ and selling it at date $i$ incurs impact on both trades. Under the no-arbitrage condition in Section~\ref{subsec:calibration-na}, $F_i^a$ is constant in $i$ and equals $x_0^a$ when the initial spot vector $x_0$ is known.

A portfolio of $\alpha$ units of cash, $b$ forwards and $c_j$ calls at strikes $K_j$ pays $\varphi(X_i^a)$, where
$$\varphi(x)=\alpha+b(x-F_i^a)+\sum_{j=1}^J c_j(x-K_j)^+,\qquad x\in\R.$$
Here $J\geq0$ is an integer and all holdings and strikes are real. We write $\mathcal V_i^a$ for the set of such functions. The initial cost of the portfolio is
$$\alpha+\sum_{j=1}^J c_j C_i^a(K_j)=\int_{\R}\varphi(x)\,\mu_i^a(dx).$$

Every $\varphi\in\mathcal V_i^a$ is continuous and has at most linear growth. For the dual problem we allow static positions in $\Lambda:=\prod_{i=0}^N\prod_{a=1}^d C_1(\R)$. With payments held in cash until date $N$, the terminal payoff and initial cost of $\varphi=(\varphi_i^a)_{i,a}\in\Lambda$ are, respectively,
$$
  \sum_{i=0}^N\sum_{a=1}^d\varphi_i^a(X_i^a)
  \quad\text{and}\quad
  \sum_{i=0}^N\sum_{a=1}^d\int_{\R}\varphi_i^a\,d\mu_i^a.
$$

A \emph{dynamic strategy} specifies the holdings $H=(H_i)_{i=0}^{N-1}$, where $H_i=h_i(X_0,\ldots,X_i)$ for a Borel map $h_i:E^{i+1}\to E$. Its component $H_i^a$ is the number of shares of asset $a$ held from date $i$ to $i+1$. We require zero inventory before the first trade and after the last, and thus $H_{-1}=H_N=0$. The order at date $i$ is $\Delta H_i:=H_i-H_{i-1}$.

For $0\leq i\leq N$, let $K_i:=\operatorname{diag}(\kappa_i^1,\ldots,\kappa_i^d)$, where the impact coefficients $\kappa_i^a>0$ are constants, and write $\kappa=(\kappa_i^a)_{0\leq i\leq N,\,1\leq a\leq d}$. At date $i$, a signed order $q\in E$ has cash expenditure $X_i\cdot q+\tfrac12\langle q,K_iq\rangle$. This quadratic cost corresponds to linear temporary impact, as in \cite[Sections~1.3--1.5]{AC01} and \cite[Section~2.1, Remark~3]{GueantPu17}. The hedger finances all trades through a cash account that permits borrowing and lending at zero interest. Starting from $B_{-1}=0$, the self-financing balance after the trade at date $i$ is
\begin{equation}\label{eq:self-financing}
 B_i=B_{i-1}-\drev{X_i\cdot\Delta H_i}-\drev{\frac12\langle\Delta H_i,K_i\Delta H_i\rangle},
 \qquad 0\leq i\leq N.
\end{equation}
Summing \eqref{eq:self-financing} and applying summation by parts gives
the terminal trading gain
\begin{equation}\label{eq:gain}
 G_\kappa(H):=B_N=\sum_{i=0}^{N-1}\drev{H_i\cdot(X_{i+1}-X_i)}-\frac12\sum_{i=0}^N\drev{\langle H_i-H_{i-1},K_i(H_i-H_{i-1})\rangle}.
\end{equation}

We use a purely quadratic cost for each order in \eqref{eq:gain} to study how the size of a claim position affects its superhedging price per unit. Quadratic rebalancing costs also appear in the binomial superhedging model of Dolinsky and Soner~\cite[Eq.~(2.4) and Example~2.2(a)]{DS13} and, in matrix form, in the multi-asset trading model of G\^arleanu and Pedersen~\cite[Eq.~(3)]{GP13}. In continuous time, Bank, Soner and Vo{\ss}~\cite[Section~2, Eq.~(2)]{BSV17} model temporary impact using the integral of the squared trading rate. For an order of $q$ shares executed at a constant rate over a fixed window of length $\tau$, this integral equals $q^2/\tau$.

We call $H$ \emph{admissible} if its terminal loss is covered by a static position of at most linear growth, that is, if there is $c\geq0$ such that
\begin{equation}\label{eq:admissible}
 G_\kappa(H)\geq\drev{-c\Bigl(1+\sum_{i=0}^N\sum_{a=1}^d|X_i^a|\Bigr)}\qquad\text{on }\drev{\Omega}.
\end{equation}
We write $\Hh$ for this class, which depends on $\kappa$ through the trading gain.  For a claim $\Phi:\Omega\to\R$ of at most linear growth, the superhedging inequality itself implies admissibility. Indeed, if
$\varphi\in\Lambda$ and
$\sum_{i=0}^N\sum_{a=1}^d\varphi_i^a(X_i^a)+G_\kappa(H)\geq\Phi$
on $\Omega$, the growth bounds imply, for some constant $C\geq0$,
$$G_\kappa(H)\geq\Phi-\sum_{i=0}^N\sum_{a=1}^d\varphi_i^a(X_i^a) \geq-C\Bigl(1+\sum_{i=0}^N\sum_{a=1}^d|X_i^a|\Bigr).$$

Let $\Hh_b$ denote the class of strategies whose holding functions $h_i:E^{i+1}\to E$, $0\leq i<N$, are bounded and continuous. For such strategies the orders are bounded, and the trading gain in \eqref{eq:gain} has at most linear growth. Hence $\Hh_b\subseteq\Hh$ by \eqref{eq:admissible}.

\subsection{Superhedging and the primal problem}\label{subsec:superhedging}

We consider a claim paying $\Phi(X)$ at date $N$, where $\Phi:\Omega\to\mathbb R$ is upper semicontinuous and has at most linear growth. Its semi-static superhedges form the set
$$\mathcal D_\kappa(\Phi):=\Bigl\{(\varphi,H)\in\Lambda\times\Hh: \sum_{i=0}^N\sum_{a=1}^d\varphi_i^a(X_i^a)+G_\kappa(H)\geq\Phi \,\, \text{on }\Omega\Bigr\}.$$
The dual, or superhedging, value is
\begin{equation}\label{eq:dual}
  D_\kappa(\Phi):=\inf_{(\varphi,H)\in\mathcal D_\kappa(\Phi)}
  \drev{\sum_{i=0}^N\sum_{a=1}^d\int\varphi_i^a\,d\mu_i^a}.
\end{equation}

For $\mu=(\mu_i^a)_{i,a}$, the set of path laws with scalar marginals $\mu_i^a$ is
$$\Pi(\mu):=\left\{\Q\in\mathcal P(\Omega): \Q\circ(X_i^a)^{-1}=\mu_i^a,\quad 0\leq i\leq N,\,\, 1\leq a\leq d\right\}.$$

Under any $\Q\in\Pi(\mu)$, the expected payoff of a static position equals its initial cost by marginal calibration. In a superhedge this position must cover $\Phi-G_\kappa(H)$. To bound its cost from below independently of $H$, we take the supremum of the expected net trading gain over admissible strategies. For every $\Q\in\mathcal P_1(\Omega)$, \eqref{eq:admissible} ensures that $G_\kappa(H)^-\in L^1(\Q)$ for each $H\in\Hh$. We define the penalty by
\begin{equation}\label{eq:penalty}
  I_\kappa(\Q):=\sup_{H\in\Hh}\E_\Q[G_\kappa(H)]\in[0,\infty].
\end{equation}

For every $(\varphi,H)\in\mathcal D_\kappa(\Phi)$ and every $\Q\in\Pi(\mu)$, integrating the superhedging inequality and then using \eqref{eq:penalty} yields
$$
  \sum_{i=0}^N\sum_{a=1}^d\int_{\R}\varphi_i^a\,d\mu_i^a
  \geq \E_\Q[\Phi]-\E_\Q[G_\kappa(H)]
  \geq \E_\Q[\Phi]-I_\kappa(\Q).
$$
The supremum of these lower bounds is the primal value
\begin{equation}\label{eq:primal}
  P_\kappa(\Phi):=\sup_{\Q\in\Pi(\mu)}
  \bigl\{\E_\Q[\Phi]-I_\kappa(\Q)\bigr\}.
\end{equation}
Taking the infimum over superhedges on the left proves the weak duality bound $P_\kappa(\Phi)\leq D_\kappa(\Phi)$.

In frictionless MOT the gain is linear in the holdings, and the supremum of expected gains over bounded adapted strategies is finite precisely under martingale laws. In \eqref{eq:gain}, trading also incurs a cost quadratic in the orders $\Delta H_i$.

\begin{remark}[Calibration across assets]
At each quoted maturity $i$, calls on asset $a$ at all strikes determine the law $\mu_i^a$ of $X_i^a$, as in \cite[Section~2]{EGLO20} and \cite[Section~2]{BHN26}. The main formulation in \cite[Section~4]{ADOT19} and the model in \cite[Section~1.1]{GO19} fix the joint law of $X_i$ at each calibration date. The alternative in \cite[Remark~4.3]{ADOT19} fixes only the individual marginal laws. For $d\geq2$, the joint law determines the expectation of the basket call $(X_i^1+X_i^2-K)^+$. In our market this expectation can vary over $\Pi(\mu)$ with the dependence between the two assets.
\end{remark}

\subsection{Duality}\label{subsec:duality}

The first main result shows that there is no duality gap.

\begin{theorem}\label{thm:main}
Let $\drev{\mu_i^a\in\mathcal P_1(\R)}$ for $0\leq i\leq N$ \drev{and $1\leq a\leq d$}, let $\drev{\kappa_i^a>0}$, and let $\Phi:\drev{\Omega}\to\R$ be upper semicontinuous and of at most linear growth.
\begin{enumerate}
\item[\textup{(i)}]
We have
\[
 P_\kappa(\Phi)=D_\kappa(\Phi)\in[-\infty,\infty).
\]
The value $D_\kappa(\Phi)$ is unchanged when $\Lambda$ and $\Hh$ in \eqref{eq:dual} are replaced by $\drev{\prod_{i=0}^N\prod_{a=1}^d\mathcal V_i^a}$ and $\Hh_b$, respectively. Moreover, $P_\kappa(\Phi)>-\infty$ if and only if there exists $\Q\in\Pi(\mu)$ such that
\begin{equation}\label{eq:finite-drift}
 \E_\Q[X_{i+1}-X_i\mid\Fc_i]\in \drev{L^2(\Q;\R^d)},
 \qquad 0\leq i<N.
\end{equation}
In this case, the supremum in \eqref{eq:primal} is attained.

\item[\textup{(ii)}]
For every $\Q\in\mathcal P_2(\drev{\Omega})$,
\begin{equation}\label{eq:penalty-representation}
 I_\kappa(\Q)
 =\min_{M\in\mathcal M_2(\Q)}
   \frac12\E_\Q\sum_{i=0}^N
       \drev{\langle M_i-X_i,K_i^{-1}(M_i-X_i)\rangle},
\end{equation}
where $\mathcal M_2(\Q)$ is the set of \drev{$\R^d$-valued} square-integrable $(\Fc_i)$-martingales under $\Q$. The minimizer is unique. Moreover, for every $\Q\in\mathcal P_1(\drev{\Omega})$, $I_\kappa(\Q)=0$ if and only if $X$ is a $\Q$-martingale.
\end{enumerate}
\end{theorem}
Under quadratic temporary impact, finite static portfolios and bounded continuous holdings suffice to recover the superhedging value, as in frictionless MOT \cite[Section~1.5]{BHP13}.

\subsubsection{A heuristic proof of the duality}\label{subsec:heuristic}

Fix $H\in\Hh_b$. The static positions must satisfy
$\sum_{i,a}\varphi_i^a(X_i^a)\geq\Phi-G_\kappa(H)$ on $\Omega$.
Write $\pi(f)$ for the static superhedging cost of $f$.
By \eqref{eq:gain}, subtracting the trading gain adds the quadratic impact cost to the residual payoff, making \(\Phi-G_\kappa(H)\) convex in \(H\).

Starting from weak duality, we obtain
\begin{align*}
  P_\kappa(\Phi)\leq D_\kappa(\Phi)
  &\leq \inf_{H\in\Hh_b}\pi(\Phi-G_\kappa(H))\\
  &\overset{\text{static duality}}{=}
      \inf_{H\in\Hh_b}\sup_{\Q\in\Pi(\mu)}
      \int_\Omega \bigl(\Phi-G_\kappa(H)\bigr)\,d\Q\\
  &\overset{\text{minimax}}{=}
      \sup_{\Q\in\Pi(\mu)}\inf_{H\in\Hh_b}
      \E_\Q[\Phi-G_\kappa(H)]\\
  &= \sup_{\Q\in\Pi(\mu)}
      \Bigg\{\E_\Q[\Phi]-
      \underbrace{\sup_{H\in\Hh_b}\E_\Q[G_\kappa(H)]}_{I_\kappa(\Q)}
      \Bigg\}
   = P_\kappa(\Phi).
\end{align*}
Section~\ref{sec:proof} justifies the minimax interchange and the
penalty identity~\eqref{eq:fixed-restriction}.

\begin{remark}[Quadratic impact with a term linear in order size]
\label{rem:mixed-trading-costs}
Some models of trade execution combine a quadratic impact cost with a charge proportional to the number of shares traded. For example, Almgren and Chriss~\cite[Section~1.5, Eq.~(7)]{AC01} model the temporary cost of trading $q$ shares over a fixed window of length $\tau$ as $\epsilon |q| + \eta q^2/\tau$. Suppose that trading $q$ shares of asset $a$ at date $i$ incurs an additional charge $s_i^a|q|$, with deterministic coefficients $\kappa_i^a>0$ and $0\le s_i^a\le\bar s<\infty$. The net trading gain is
$$G_{\kappa,s}(H) := G_\kappa(H)-\sum_{i=0}^N\sum_{a=1}^d s_i^a|\Delta H_i^a|.$$
Let $\Hh_{\kappa,s}$ consist of the dynamic strategies satisfying \eqref{eq:admissible} with $G_{\kappa,s}$ in place of $G_\kappa$. The superhedging value $D_{\kappa,s}(\Phi)$ is the infimum of the static cost in \eqref{eq:dual} over all $(\varphi,H)\in\Lambda\times\Hh_{\kappa,s}$ such that
$$\sum_{i=0}^N\sum_{a=1}^d\varphi_i^a(X_i^a)+G_{\kappa,s}(H) \ge \Phi \qquad\text{on }\Omega.$$
Under the assumptions of Theorem~\ref{thm:main},
$$D_{\kappa,s}(\Phi)=P_{\kappa,s}(\Phi) :=\sup_{\Q\in\Pi(\mu)}\bigl\{\E_\Q[\Phi]-I_{\kappa,s}(\Q)\bigr\},$$
where $I_{\kappa,s}(\Q) := \sup_{H\in\Hh_{\kappa,s}}\E_\Q[G_{\kappa,s}(H)]$.

Indeed, fix $H\in\Hh_{\kappa,s}$ with admissibility constant $c$, a history $x_{0:i}$ with $i<N$, and an asset $a$. For $t>0$, consider two continuations. At date $i+1$ only the price of asset $a$ changes, by $+t$ on one path and by $-t$ on the other. All prices then remain fixed through date $N$. The smaller of the two remaining gains before costs is $-t|H_i^a(x_{0:i})|$. On either path the sum of absolute prices at each of the $N-i$ remaining dates is at most $\|x_i\|_1+t$. Since trading costs are nonnegative, the admissibility bound implies
$$ t|H_i^a(x_{0:i})| \leq (N-i)ct + \sum_{j=0}^{i-1}H_j(x_{0:j})\cdot(x_{j+1}-x_j) +c\Bigl(1+\sum_{j=0}^{i}\|x_j\|_1+(N-i)\|x_i\|_1\Bigr).$$
Dividing by $t$ and letting $t\to\infty$ yields $|H_i^a(x_{0:i})|\leq (N-i)c$ for every history.

Fix $\Q\in\Pi(\mu)$. As in the proof of Proposition~\ref{prop:fixed}, approximate the holdings in $L^2(\Q)$ by continuous functions of their histories, clipped at the above bounds. Along a subsequence, the holdings converge $\Q$-almost surely, and the gains converge in expectation by their common linear growth bound. Hence $\Hh_b$ suffices to compute $I_{\kappa,s}(\Q)$. The proportional fee is convex and Lipschitz in the orders and, for $H\in\Hh_b$, bounded and continuous on $\Omega$. The static duality and minimax steps in the proof of Theorem~\ref{thm:main}\textup{(i)} now apply.

\end{remark}

\begin{remark}[Adapted quadratic impact]
In the purely quadratic model, let each coefficient $\kappa_i^a:E^{i+1}\to(0,\infty)$ be a continuous function of the price history $x_{0:i}$ observed when the order is placed. Suppose $0<\underline\kappa\leq\kappa_i^a(x_{0:i})\leq\bar\kappa<\infty$ uniformly over dates, assets and histories. Use the diagonal matrices $K_i(X_{0:i})$ with these entries in \eqref{eq:gain}.  Under the marginal and payoff assumptions of Theorem~\ref{thm:main}, the resulting values satisfy $P_\kappa(\Phi)=D_\kappa(\Phi)$. The holding bound in Remark~\ref{rem:mixed-trading-costs} applies because the trading costs are nonnegative. The uniform upper bound on the coefficients permits the same dominated convergence argument for the penalty. For $H\in\Hh_b$ the impact cost is bounded and continuous on $\Omega$, and on each fixed path it is convex in the orders. The static duality and minimax steps in the proof of Theorem~\ref{thm:main}\textup{(i)} therefore apply.
\end{remark}

\begin{remark}\label{rem:subhedging}
Under the marginal and impact assumptions of Theorem~\ref{thm:main}, let $\Phi:\drev{\Omega}\to\R$ be Borel measurable and of at most linear growth. The buyer maximizes the proceeds from selling a static position whose liability is covered by the claim and dynamic trading.
The corresponding subhedging value is
$$
 \underline D_\kappa(\Phi)
 :=
 \sup\left\{
 \drev{\sum_{i=0}^{N}\sum_{a=1}^d\int\varphi_i^a\,d\mu_i^a}:
 (\varphi,H)\in\Lambda\times\Hh,
  \drev{\sum_{i=0}^{N}\sum_{a=1}^d\varphi_i^a(X_i^a)}
  \leq\Phi+G_\kappa(H)\ \text{on }\drev{\Omega}
 \right\}.
$$
If $\Phi$ is lower semicontinuous, the substitution $(\varphi,H)\mapsto(-\varphi,H)$ and Theorem~\ref{thm:main}\textup{(i)}, applied to $-\Phi$, imply
$$\underline D_\kappa(\Phi) = -D_\kappa(-\Phi) = \inf_{\Q\in\Pi(\mu)} \left\{\E_\Q[\Phi]+I_\kappa(\Q)\right\}.$$
The value is finite if and only if $I_\kappa(\Q)<\infty$ for some $\Q\in\Pi(\mu)$; in this case, the infimum is attained. Under these assumptions, the reductions to finite static portfolios at dates $0,\ldots,N$ and to dynamic strategies in $\Hh_b$ also hold.
\end{remark}

\subsubsection{Dual attainment}\label{subsec:dual-attainment}

Under compact support and continuity of the claim, the dual
infimum is attained as well.

\begin{proposition}[Dual attainment]
\label{prop:dual-attainment}
Let $N\geq1$, \drev{$d\geq1$,} let $\drev{\kappa_i^a>0}$, and suppose
that \drev{all $\mu_i^a$, $0\leq i\leq N$, $1\leq a\leq d$,} are compactly supported probability
measures on $\drev{\R}$.
If $\Phi:\drev{\Omega}\to\R$ is continuous and of at most linear growth,
then there exists $(\varphi^*,H^*)\in\mathcal D_\kappa(\Phi)$ such that
\[
 \drev{\sum_{i=0}^N\sum_{a=1}^d\mu_i^a(\varphi_i^{a,*})}=D_\kappa(\Phi).
\]
\end{proposition}

The proof  uses dynamic programming with the previous holding as an additional state variable; compare \cite[Theorems~3.1 and~4.5]{LV23} and \cite[Proposition~5.6 and Section~5.2]{NS18}.

\subsection{Cross-maturity no-arbitrage}
\label{subsec:calibration-na}

For $\Phi=\sum_{i,a}\varphi_i^a(X_i^a)$ with $\varphi\in\Lambda$, marginal calibration and Theorem~\ref{thm:main}\textup{(i)} imply
$$D_\kappa(\Phi)=P_\kappa(\Phi) = \sum_{i,a}\int_{\R}\varphi_i^a\,d\mu_i^a+D_\kappa(0).$$

A semi-static portfolio with initial cost $-c<0$ and nonnegative terminal payoff on every path secures an initial gain of $c$. We denote $\mathrm{NA}_\kappa$ for the absence of such a portfolio. Since the zero portfolio is feasible, $\mathrm{NA}_\kappa$ holds exactly when $D_\kappa(0)=0$. Evaluating \eqref{eq:primal} at the zero payoff yields
$$-D_\kappa(0)=-P_\kappa(0)=\inf_{\Q\in\Pi(\mu)}I_\kappa(\Q).$$
Let $\mathcal M(\mu)$ denote the set of $\Q\in\Pi(\mu)$ under which $X$ is an $(\Fc_i)$-martingale.

\begin{proposition}\label{prop:convex-order-na}
Under the marginal and impact assumptions of Theorem~\ref{thm:main},
$$
  \mathrm{NA}_\kappa
  \quad\Longleftrightarrow\quad \mathcal M(\mu)\ne\varnothing
  \quad\Longleftrightarrow\quad
  \mu_0^a\cv\mu_1^a\cv\cdots\cv\mu_N^a
  \quad\text{for every }a.
$$
\end{proposition}

\section{The one-period problem and its weak transport formulation}
\label{sec:one-period}

For one asset over one trading period, the pricing problem of Section~\ref{sec:main} admits a weak transport formulation. Under each coupling of the two marginal laws, the impact penalty is proportional to the expected squared difference between $X_0$ and the conditional mean of $X_1$ given $X_0$, as in the quadratic barycentric transport problem studied in \cite{GJ18}.

For $i=0,1$, write $\mu_i:=\mu_i^1\in\mathcal P_1(\R)$ and $\kappa_i:=\kappa_i^1>0$. The payoff $\Phi:\R^2\to\R$ is upper semicontinuous and has at most linear growth. A position $H_0=h_0(X_0)$, with $h_0:\R\to\R$ Borel, opens with an order of $h_0(X_0)$ shares and closes with an order of $-h_0(X_0)$ shares. Both orders have the same absolute size, and \eqref{eq:gain} becomes
$$G_\kappa(h_0)=h_0(X_0)(X_1-X_0)-\frac{\kappa_0+\kappa_1}{2}h_0(X_0)^2.$$

The calibrated laws are the couplings $\gamma\in\Pi(\mu_0,\mu_1)$. Write $\gamma(dx,dy)=\mu_0(dx)\gamma_x(dy)$. Since $\mu_1\in\mathcal P_1(\R)$, we may take the Borel kernel $\gamma_x$ with values in $\mathcal P_1(\R)$ by setting $\gamma_x=\delta_0$ on a $\mu_0$-null Borel set if necessary. The conditional drift of the price under $\gamma$ is
$$b_\gamma(x):=\int_{\R} y\,\gamma_x(dy)-x,\qquad \mu_0\text{-a.e. }x\in\R.$$

\subsection{Weak transport formulation}
\label{subsec:one-period-formulation}

\begin{proposition}\label{prop:one-period-formulation}
For every $\gamma\in\Pi(\mu_0,\mu_1)$,
\begin{equation}\label{eq:wot-barycentric}
 I_\kappa(\gamma)=\frac1{2(\kappa_0 + \kappa_1)}\int_{\borelrev{\R}}b_\gamma(x)^2\,\mu_0(dx)\in[0,\infty].
\end{equation}
Consequently, the primal and dual values admit the following weak transport representation:
\begin{equation}\label{eq:one-period-cost}
 P_\kappa(\Phi)=D_\kappa(\Phi)=-\inf_{\gamma\in\Pi(\mu_0,\mu_1)}\int_{\borelrev{\R}}C^\Phi_\kappa(x,\gamma_x)\,\mu_0(dx),
\end{equation}
where, for $x\in \borelrev{\R}$ and $p\in\mathcal P_1(\borelrev{\R})$,
$$
 C^\Phi_\kappa(x,p):=\frac1{2(\kappa_0 + \kappa_1)}\Bigl(x-\int_{\borelrev{\R}} y\,p(dy)\Bigr)^2-\int_{\borelrev{\R}}\Phi(x,y)\,p(dy).
$$
The value in \eqref{eq:one-period-cost} is finite if and only if $b_\gamma\in L^2(\mu_0)$ for some $\gamma\in\Pi(\mu_0,\mu_1)$. In this case, the infimum is attained.
\end{proposition}

The first term in $C_\kappa^\Phi(x,p)$ is $1/[2(\kappa_0+\kappa_1)]$ times the quadratic barycentric cost in \cite[Eq.~(1.5)]{BBP19}. Using the disintegration $\gamma(dx,dy)=\mu_0(dx)\gamma_x(dy)$, we can write \eqref{eq:one-period-cost} as
\begin{equation}\label{eq:one-period-value}
 P_\kappa(\Phi)
 =\sup_{\gamma\in\Pi(\mu_0,\mu_1)}
   \left\{
     \int_{\R^2}\Phi\,d\gamma
     -\frac{1}{2(\kappa_0+\kappa_1)}\int_{\R} b_\gamma^2\,d\mu_0
   \right\},
\end{equation}
with objective value $-\infty$ when $b_\gamma\notin L^2(\mu_0)$.

\subsection{Structure of optimal couplings}
\label{subsec:one-period-structure}

In the following theorem, we show that whenever the value in \eqref{eq:one-period-value} is finite, all maximizing couplings have the same conditional drift $\mu_0$-almost everywhere. Consequently, the conditional mean $Z=\E_\gamma[X_1\mid X_0]=X_0+b_\gamma(X_0)$ has the same law under every maximizing coupling $\gamma$.

This choice of $Z$ satisfies $\E_\gamma[X_1\mid X_0,Z]=Z$. To characterize its law under finite second moments, we allow $Z$ to be random given $X_0$ subject to the same condition. Even when the law of $Z$ is fixed, the expectation of $\Phi(X_0,X_1)$ can vary with the coupling of the endpoints. Let $(X_0,Z,X_1)$ denote the coordinates on $\R^3$. For $\mu_0,\mu_1,\zeta\in\mathcal P_2(\R)$, set
\begin{equation}\label{eq:one-period-lift-set}
  \mathcal A(\zeta):=\bigl\{\Gamma\in\mathcal P_2(\R^3):
    \Gamma\circ X_0^{-1}=\mu_0,\,\,
    \Gamma\circ Z^{-1}=\zeta,\,\,
    \Gamma\circ X_1^{-1}=\mu_1,\,\,
    \E_\Gamma[X_1\mid X_0,Z]=Z\bigr\}.
\end{equation}

\begin{theorem}\label{thm:one-period-structure}
\begin{enumerate}
\item[\textup{(i)}]
Suppose that $P_\kappa(\Phi)>-\infty$. There exists a unique $b_\kappa^*\in L^2(\mu_0)$ such that $b_\gamma=b_\kappa^*$ $\mu_0$-a.s.\ for every maximizer $\gamma$ of \eqref{eq:one-period-value}. Moreover, for every $\gamma\in\Pi(\mu_0,\mu_1)$,
\begin{equation}\label{eq:one-period-gap}
 \frac{1}{2(\kappa_0+\kappa_1)}
 \int_{\borelrev{\R}}|b_\gamma-b_\kappa^*|^2\,d\mu_0
 \leq
 P_\kappa(\Phi)
 +\int_{\borelrev{\R}}C_\kappa^\Phi(x,\gamma_x)\,\mu_0(dx).
\end{equation}

\item[\textup{(ii)}]
Suppose that $\mu_i\in\mathcal P_2(\borelrev{\R})$ for $i=0,1$. For $\zeta\in\mathcal P_2(\borelrev{\R})$, $\mathcal A(\zeta)\neq\varnothing$ if and only if $\zeta\cv\mu_1$. Whenever $\zeta\cv\mu_1$, every coupling $\rho\in\Pi(\mu_0,\zeta)$ admits an extension $\Gamma\in\mathcal A(\zeta)$ with $(X_0,Z)$-marginal $\rho$. Moreover,
\begin{equation}\label{eq:one-period-lift}
 P_\kappa(\Phi)
 =
 \sup_{\zeta\cv\mu_1}\ \sup_{\Gamma\in\mathcal A(\zeta)}
 \E_\Gamma\left[
   \Phi(X_0,X_1)
   -\frac{|X_0-Z|^2}{2(\kappa_0+\kappa_1)}
 \right].
\end{equation}
The unique maximizer of the outer supremum is $\zeta^*:=(\mathrm{id}+b_\kappa^*)_\#\mu_0$. For this intermediate marginal, the inner supremum is attained, and its maximizers are precisely the laws of $(X_0,X_0+b_\kappa^*(X_0),X_1)$ under maximizers $\gamma$ of \eqref{eq:one-period-value}.

\item[\textup{(iii)}]
Suppose that $\mu_i\in\mathcal P_2(\borelrev{\R})$ for $i=0,1$ and $\Phi=0$. Then
\begin{equation}\label{eq:one-period-projection}
 P_\kappa(0)=D_\kappa(0)
 =-\frac{1}{2(\kappa_0+\kappa_1)}
   \inf_{\zeta\cv\mu_1}\Wc_2(\mu_0,\zeta)^2.
\end{equation}
The infimum is attained at a unique measure $\mu_0^*\in\mathcal P_2(\borelrev{\R})$. For every $\kappa\in(0,\infty)^2$, $\mu_0^*=(\mathrm{id}+b_\kappa^*)_\#\mu_0$, and the map $\mathrm{id}+b_\kappa^*$ agrees $\mu_0$-a.s.\ with a nondecreasing $1$-Lipschitz function on $\R$. The identity $\mu_0^*=\mu_0$ holds if and only if $\mu_0\cv\mu_1$.
\end{enumerate}
\end{theorem}

For one asset over one period with square-integrable marginals, part~\textup{(iii)} quantifies the no-arbitrage criterion of Proposition~\ref{prop:convex-order-na}. Among semi-static portfolios with nonnegative terminal payoff on every path, the supremum of the net cash receipts at inception is
$$-D_\kappa(0) = \inf_{\gamma\in\Pi(\mu_0,\mu_1)} I_\kappa(\gamma) = \frac{\Wc_2(\mu_0,\mu_0^*)^2}{2(\kappa_0+\kappa_1)}.$$
This supremum vanishes exactly when $\mu_0\cv\mu_1$. For fixed marginals that violate convex order, this supremum is a positive constant times $(\kappa_0+\kappa_1)^{-1}$ and therefore tends to $+\infty$ as $\kappa_0+\kappa_1\downarrow0$.

\keeprevisioncolorstrue%
\section{OT and MOT limits under quadratic price impact}\label{subsec:asymptotics}

Fix the marginal laws $\mu_i^a\in\mathcal P_1(\R)$ and the impact coefficients $\kappa_i^a>0$ for $0\leq i\leq N$ and $1\leq a\leq d$. We replace $\kappa$ by $\varepsilon\kappa$, with $\varepsilon>0$, while keeping the marginals fixed. An order of $\Delta H_i^a$ shares then incurs an impact cost of $\varepsilon\kappa_i^a(\Delta H_i^a)^2/2$. Throughout this section $\Phi:\Omega\to\R$ is upper semicontinuous and of at most linear growth.

\subsection{Limits of prices and optimal laws}

\begin{proposition}\label{prop:asy-calculus}
\begin{enumerate}
\item[\textup{(i)}]
For every $\Q\in\mathcal P_1(\Omega)$ and $\varepsilon>0$,
\begin{equation}\label{eq:asy-scale}
 I_{\varepsilon\kappa}(\Q)
 =\frac{I_\kappa(\Q)}{\varepsilon},
\end{equation}
where the equality also holds when $I_\kappa(\Q)=+\infty$.

\item[\textup{(ii)}]
For $\Q\in\mathcal P_1(\Omega)$, set $b_i=\E_\Q[X_{i+1}-X_i\mid\Fc_i]$, $0\leq i<N$. If $b_i\in L^2(\Q;\R^d)$ for every $i<N$, then
\begin{equation}\label{eq:asy-drift}
 \frac{1}{8\max_{j,a}\kappa_j^a}
       \sum_{i=0}^{N-1}\E_\Q[|b_i|^2]
 \leq I_\kappa(\Q)
 \leq\frac{N^2}{2\min_{j,a}\kappa_j^a}
       \sum_{i=0}^{N-1}\E_\Q[|b_i|^2].
\end{equation}

\item[\textup{(iii)}]
The map $\Q\mapsto I_\kappa(\Q)$ is convex and lower semicontinuous on $\Pi(\mu)$ for the weak topology. For every $\varepsilon>0$,
\[
 P_{\varepsilon\kappa}(\Phi)=D_{\varepsilon\kappa}(\Phi)
 =\sup_{\Q\in\Pi(\mu)}
    \left\{\E_\Q[\Phi]-\frac{I_\kappa(\Q)}{\varepsilon}\right\},
\]
and $\varepsilon\mapsto P_{\varepsilon\kappa}(\Phi)$ is nondecreasing with values in $[-\infty, \infty)$. If $I_\kappa(\Q)<\infty$ for some $\Q\in\Pi(\mu)$, the value is finite and the supremum is attained for every $\varepsilon>0$.
\end{enumerate}
\end{proposition}

We next compare $P_{\varepsilon\kappa}(\Phi)$ with the ordinary and martingale transport values
\begin{equation}\label{eq:asy-endpoints}
 O:=\max_{\Q\in\Pi(\mu)}\E_\Q[\Phi],\qquad
 m:=\max_{\Q\in\mathcal M(\mu)}\E_\Q[\Phi],
\end{equation}
where $m$ is defined when $\mathcal M(\mu)\ne\varnothing$, equivalently when $\mu_0^a\cv\mu_1^a\cv\cdots\cv\mu_N^a$ for every asset $a$.

\begin{theorem}\label{thm:asy-endpoints}
Suppose that $\mu_i^a\in\mathcal P_2(\R)$ for every $i,a$. For each $\varepsilon>0$, let $\Q_\varepsilon$ be any optimizer in Proposition~\ref{prop:asy-calculus}\textup{(iii)}.
\begin{enumerate}
\item[\textup{(i)}] As $\varepsilon\to\infty$,
\begin{equation}\label{eq:asy-OT}
 P_{\varepsilon\kappa}(\Phi)\uparrow O.
\end{equation}
Every weak subsequential limit $\Q_\infty$ of $\Q_\varepsilon$ as $\varepsilon\to\infty$ satisfies $\E_{\Q_\infty}[\Phi]=O$ and minimizes $I_\kappa$ among laws in $\Pi(\mu)$ with expected payoff $O$.

\item[\textup{(ii)}] As $\varepsilon\downarrow0$,
\begin{equation}\label{eq:asy-MOT}
 P_{\varepsilon\kappa}(\Phi)\downarrow
 \begin{cases}
 m,&\mathcal M(\mu)\ne\varnothing,\\
 -\infty,&\mathcal M(\mu)=\varnothing
 \end{cases}.
\end{equation}
Every weak subsequential limit $\Q_0$ of $\Q_\varepsilon$ as $\varepsilon\downarrow0$ minimizes $I_\kappa$ on $\Pi(\mu)$ and maximizes expected payoff among these minimizers.
\end{enumerate}
\end{theorem}

\begin{remark}\label{rem:asy-P1}
Notice that \eqref{eq:asy-MOT} holds under first moments alone. If $I_\kappa(\Q)<\infty$ for some $\Q\in\Pi(\mu)$, then $I_\kappa$ attains a finite minimum on $\Pi(\mu)$ by weak compactness and lower semicontinuity. The comparison argument in the proof of Theorem~\ref{thm:asy-endpoints}\textup{(ii)}, starting from \eqref{eq:asy-small-comparison}, then applies. If $I_\kappa\equiv+\infty$ on $\Pi(\mu)$, then $\mathcal M(\mu)=\varnothing$ and $P_{\varepsilon\kappa}(\Phi)=-\infty$ for every $\varepsilon>0$.

The OT price limit \eqref{eq:asy-OT} can fail under the first moment assumption alone. Let $d=1$, $N=2$, $\Phi=0$, and take
$$\mu_0=\mu_2=\delta_0,\qquad \mu_1(dx)=2(1+x)^{-3}\mathbf{1}_{\{x\geq0\}}\,dx.$$
The law $\mu_1$ has finite first moment and infinite second moment. Under the unique $\Q\in\Pi(\mu)$, the conditional drift at time $1$ is $\E_\Q[X_2-X_1\mid\Fc_1]=-X_1\notin L^2(\Q)$. By Proposition~\ref{prop:fixed}, $I_{\varepsilon\kappa}(\Q)=+\infty$ for every $\varepsilon>0$. Consequently, $P_{\varepsilon\kappa}(0)=-\infty$ for every $\varepsilon>0$, whereas $O=0$.
\end{remark}

\subsection{Convergence rates}

We first establish an upper bound of order $1/\varepsilon$ on the gap between the penalized price $P_{\varepsilon\kappa}(\Phi)$ and the OT value $O$.

\begin{proposition}\label{prop:rate-OT}
Under the assumptions of Theorem~\ref{thm:asy-endpoints}, set
\[
 j_O:=\min\{I_\kappa(\Q):\Q\in\Pi(\mu),\ \E_\Q[\Phi]=O\}.
\]
Then $j_O<\infty$ and $0\leq O-P_{\varepsilon\kappa}(\Phi)\leq\frac{j_O}{\varepsilon}$ for every $\varepsilon > 0$. Moreover,
\begin{equation}\label{eq:rate-OT-first-order}
 P_{\varepsilon\kappa}(\Phi)
   =O-\frac{j_O}{\varepsilon}
          +o(\varepsilon^{-1}),\qquad \varepsilon\to\infty.
\end{equation}
\end{proposition}

As $\varepsilon\downarrow0$, the gap $P_{\varepsilon\kappa}(\Phi)-m$ is bounded by a constant times $\varepsilon$ if $\Phi$ admits a frictionless superhedge with initial cost $m$ and bounded holdings. Executing the same strategy incurs a total impact cost of $\frac{\varepsilon}{2}\sum_{i,a}\kappa_i^a(\Delta H_i^a)^2$, and each order $\Delta H_i^a=H_i^a-H_{i-1}^a$ is a difference of bounded holdings.

\begin{proposition}\label{prop:rate-additive}
For each asset $a$, let $\mu_0^a,\ldots,\mu_N^a\in\mathcal P_1(\R)$ satisfy $\mu_0^a\cv\cdots\cv\mu_N^a$. Suppose there are static payoffs $\varphi_i^a\in L^1(\mu_i^a)$ and a bounded Borel adapted strategy $H$ such that
\begin{equation}\label{eq:rate-hedge}
 \begin{gathered}
 \Phi\leq \sum_{i=0}^N\sum_{a=1}^d\varphi_i^a(X_i^a)
       +\sum_{i=0}^{N-1}H_i\cdot(X_{i+1}-X_i)
 \quad \Q\text{-a.s. for every }\Q\in\Pi(\mu),\\
 \sum_{i=0}^N\sum_{a=1}^d\mu_i^a(\varphi_i^a)=m.
 \end{gathered}
\end{equation}
Choose $C_i^a$ such that $|H_i^a|\leq C_i^a$ for $0\leq i<N$, and set $C_{-1}^a=C_N^a=0$. Then, for every $\varepsilon>0$,
\begin{equation}\label{eq:rate-smooth}
 0\leq P_{\varepsilon\kappa}(\Phi)-m
 \leq\frac{\varepsilon}{2}
     \sum_{i=0}^N\sum_{a=1}^d\kappa_i^a(C_i^a+C_{i-1}^a)^2.
\end{equation}
\end{proposition}

\begin{remark}[Additive payoffs]\label{rem:additive-hedge}
Suppose that the marginal assumptions of Proposition~\ref{prop:rate-additive} hold and that, on $\prod_{i,a}\operatorname{supp}\mu_i^a$,
$$\Phi(x)=\sum_{a=1}^d\sum_{i=0}^{N-1}c_i^a(x_i^a,x_{i+1}^a),$$
where each $c_i^a:\R^2\to\R$ is Lipschitz. For each $i<N$ and $a$, let $J_i^a=\operatorname{conv}(\operatorname{supp}\mu_{i+1}^a)$ and suppose there is a Lipschitz function $u_i^a:J_i^a\to\R$ such that $y\mapsto-c_i^a(x,y)+u_i^a(y)$ is convex on $J_i^a$ for $\mu_i^a$-almost every $x$. Then \eqref{eq:rate-hedge} holds with
$$C_i^a=18\max\{\operatorname{Lip}(c_i^a),\operatorname{Lip}(u_i^a)\}.$$

Indeed, let $m_i^a$ be the maximum of the expected payoff $c_i^a$ over martingale couplings of $\mu_i^a$ and $\mu_{i+1}^a$. These maxima are attained by weak compactness of the martingale coupling sets and continuity of the payoff integrals. Apply \cite[Theorem~2.5 and Remark~2.6]{BLO19} to $-c_i^a$ and change sign. The resulting scalar superhedge has cost $m_i^a$ and a Borel holding $h_i^a$ bounded by $C_i^a$ on $J_i^a$. Convex order implies $\operatorname{supp}\mu_i^a\subseteq J_i^a$, hence we may set $h_i^a=0$ off $J_i^a$. If $J_i^a$ is a singleton, both marginals are the same point mass and a constant static payoff suffices.

Every $\Q\in\mathcal M(\mu)$ induces a martingale coupling for each pair $(X_i^a,X_{i+1}^a)$, hence $m\leq\sum_{i,a}m_i^a$. Conversely, for each asset use the transition kernels of scalar maximizing couplings at successive dates, then take the product of the resulting path laws across assets. Under this product law the coordinate processes are independent, hence $\E[X_{i+1}^a\mid\Fc_i]=X_i^a$ for every $i,a$. The product law attains every $m_i^a$, hence $m=\sum_{i,a}m_i^a$.

Each scalar hedge inequality holds for $\mu_i^a$-almost every $x$ and every $y\in\operatorname{supp}\mu_{i+1}^a$. Their sum therefore holds under every $\Q\in\Pi(\mu)$. Combining the static payoffs at each date and taking $H_i^a=h_i^a(X_i^a)$ produces the hedge in \eqref{eq:rate-hedge}, with total static cost $m$.
\end{remark}

\begin{example}[A two-asset spread in one period]\label{ex:multi-rate-counterexample}
Let $N=1$, $d=2$, and take $\kappa_0^a=\kappa_1^a=1$ and
$\mu_0^a=\mu_1^a=\operatorname{Unif}[0,1]$ for both assets. Consider
\[
 f(z)=|z^1-z^2|,\qquad \Phi(x,y)=f(y)-f(x).
\]
For fixed $x$, the term $|y^1-y^2|$ prevents $\Phi(x,y)$ from splitting
into separate payoffs on the two assets as in Remark~\ref{rem:additive-hedge}.
Under every $\Q\in\mathcal M(\mu)$, each coordinate has the same second
moment at both dates, hence $X_1=X_0$ $\Q$-a.s. Thus $m=0$, and the
zero hedge covers $\Phi$ under every martingale law.

Fix $0<h\leq1/4$ and let $U$ be uniform on $[0,1]$. Given $U=u$,
choose $u-h$ or $u+h$ with probability $1/2$ each. If the chosen point
lies outside $[0,1]$, reflect it at the endpoint it crossed. Denote the
result by $Y$. For the right and left choices, the respective densities
of $Y$ are $\mathbf1_{[h,1]}+\mathbf1_{[1-h,1]}$ and
$\mathbf1_{[0,1-h]}+\mathbf1_{[0,h]}$. Their average is one almost
everywhere on $[0,1]$. Thus the law $\Q_h$ of $X_0=(U,U)$ and
$X_1=(Y,U)$ belongs to $\Pi(\mu)$.

Under $\Q_h$ the payoff is $|Y-U|$. On $h\leq U\leq1-h$, both steps
have size $h$ and the conditional drift $b(U):=\E[Y-U\mid U]$ is zero.
On the two boundary intervals $|b(U)|\leq h$. Hence
\[
 \E_{\Q_h}[\Phi]\geq h(1-2h)\geq h/2,\qquad
 \E[b(U)^2]\leq 2h^3.
\]
The orders $H_0$ and $-H_0$ incur a total impact cost of
$\varepsilon|H_0|^2$. By \eqref{eq:fixed-max}, the penalty is the
maximum of $\E[H_0^1 b(U)-\varepsilon|H_0|^2]$. Completing the square,
we obtain
\[
 I_{\varepsilon\kappa}(\Q_h)
 =\frac{\E[b(U)^2]}{4\varepsilon}\leq\frac{h^3}{2\varepsilon}.
\]
For $h=\sqrt{\varepsilon}/2$, the objective at $\Q_h$ in \eqref{eq:primal}
is at least $h/2-h^3/(2\varepsilon)$. Therefore
\begin{equation}\label{eq:multi-rate-counterexample}
 P_{\varepsilon\kappa}(\Phi)-m\geq\frac{3}{16}\sqrt{\varepsilon},
 \qquad 0<\varepsilon\leq\frac14.
\end{equation}
Hence the linear upper bound in \eqref{eq:rate-smooth} can fail if
coverage is required only under martingale laws.
\end{example}

\keeprevisioncolorsfalse%

\section{Numerical examples}\label{sec:fin}

We first study a forward-start call on one asset. With the two lognormal marginals fixed, we examine how the seller's upper bound and the expected number of shares traded per claim vary with the impact coefficient. With marginals fitted to SPX option quotes, we compare the price bounds per claim across order sizes and strike ratios. Our second claim is a call on the sum of two asset prices at maturity. We keep the four marginal laws fixed and compare the seller's upper bound before and after interchanging the two impact coefficients.

\subsection{Single-asset forward-start options}\label{subsec:fin-setup}

The strike of the forward-start call is set at $kX_1$ when $X_1$ is observed. At maturity the call pays $\Phi_k(X_1,X_2)=(X_2-kX_1)^+$, as in \cite[Section~4.1]{BHP13} and \cite[Section~5.2]{BAMOT26}. The initial spot is $X_0=1$. The prices $X_1$ and $X_2$ have laws $\mu_1,\mu_2\in\mathcal P_2(\R_+)$, respectively. Both means equal the initial spot, and $\mu_1\cv\mu_2$. With these laws fixed, the expected payoff can still vary with the coupling of the two prices.

\subsubsection{Prices and holdings with lognormal marginals}\label{subsec:fin-simulation}

The strike is the spot price observed after six months ($T_1=0.5$, $k=1$), and the call expires after one year ($T_2=1$). We assume zero interest and no dividends. We fix the annualized volatilities at $\sigma_1=19.5\%$ and $\sigma_2=17.5\%$. For $i=1,2$, set $v_i=T_i\sigma_i^2$ and take $\log X_i\sim N(-v_i/2,v_i)$.

For comparison, we first price the call assuming that the log returns over $[0,T_1]$ and $[T_1,T_2]$ are independent and Gaussian. With the marginal laws above, the expected payoff is $429.70$ basis points of the initial spot (Appendix~\ref{app:fin-grid}).

To compute the seller's upper bound, let $\gamma$ range over the couplings of $\mu_1$ and $\mu_2$ and write $b_\gamma(x)=\E_\gamma[X_2-X_1\mid X_1=x]$ for its conditional drift. Since the expectations of $X_1$ and $X_2$ both equal $X_0$, the initial holding that maximizes expected net trading gains under $\gamma$ is zero. The penalty thus reduces to the one-period penalty in \eqref{eq:wot-barycentric}, with $\eta:=\kappa_1+\kappa_2$. Writing $P_\eta(\Phi_1)$ for the seller's upper bound, we obtain from \eqref{eq:primal}
\begin{equation}\label{eq:fin-wot-price}
 P_\eta(\Phi_1)
 =\sup_{\gamma\in\Pi(\mu_1,\mu_2)}
 \left\{\int (y-x)^+\,\gamma(dx,dy)
       -\frac{1}{2\eta}\int b_\gamma(x)^2\,\mu_1(dx)\right\}.
\end{equation}

\begin{figure}[H]
 \centering
 \includegraphics[width=\textwidth]{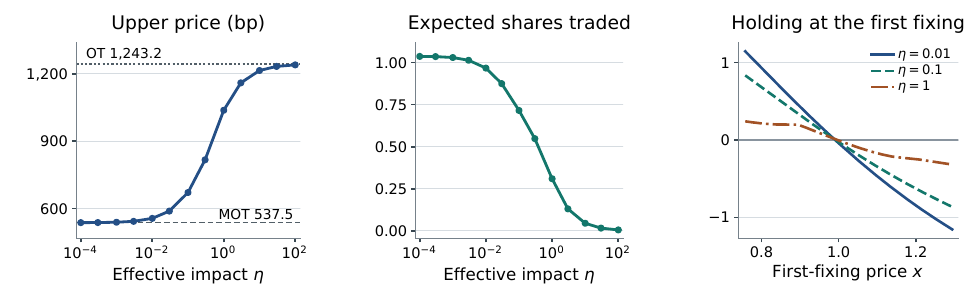}
\caption{At-the-money forward-start call with lognormal marginals. As the impact coefficient $\eta$ varies, the left panel shows the seller's upper bound in basis points of the initial spot, and the middle panel shows the expected number of shares traded per claim, including liquidation. Both panels use the full discrete marginals. The horizontal lines in the left panel mark the MOT and OT upper values. For $\eta=0.01,0.1,1$, the right panel plots the signed holding at $T_1$ against $X_1$ at all 256 support points of the first marginal.}
 \label{fig:fin-bs}
\end{figure}

We solve \eqref{eq:fin-wot-price} at thirteen positive values of $\eta$ with the same two discrete marginals, each on 256 support points (Appendix~\ref{app:fin-grid}). For these marginals, the seller's upper bound converges to the MOT upper value as $\eta\downarrow0$ and to the OT upper value as $\eta\uparrow\infty$ (Theorem~\ref{thm:asy-endpoints}). The horizontal lines in the left panel of Figure~\ref{fig:fin-bs} mark these values at $537.54$ and $1{,}243.15$ basis points, respectively. As $\eta$ increases from $10^{-4}$ to $100$ across the thirteen computed values, the weight on squared conditional drift in \eqref{eq:fin-wot-price} decreases and the seller's upper bound rises from $537.75$ to $1{,}240.08$ basis points of the initial spot.

In the middle panel of Figure~\ref{fig:fin-bs}, the expected number
of shares traded per claim, $V_\eta$, decreases at each step of the
grid, from $1.037$ at $\eta=10^{-4}$ to $0.005$ at $\eta=100$. We compute $V_\eta$ under a coupling $\gamma_\eta$ attaining the supremum in \eqref{eq:fin-wot-price}. When $X_1=x$, the holding that maximizes expected net trading gains under this coupling is
$b_{\gamma_\eta}(x)/\eta$ shares. Counting the opening trade at
$T_1$ and the closing trade at $T_2$, we obtain
\begin{equation}\label{eq:fin-single-volume}
 V_\eta=\frac{2}{\eta}\int |b_{\gamma_\eta}(x)|\,\mu_1(dx).
\end{equation}
For each fixed $\eta$, all maximizing couplings have the same
conditional drift $\mu_1$-almost everywhere and hence the same
$V_\eta$ (Theorem~\ref{thm:one-period-structure}\textup{(i)}).
In the right panel, the position in the underlying is long near
$X_1=0.80$ and short near $X_1=1.20$. At both prices, the position
is smaller in absolute value at $\eta=1$ than at $\eta=0.01$.

The right panel of Figure~\ref{fig:fin-bs} shows the holding at $T_1$. As $\eta$ increases from $0.01$ to $1$, the number of shares bought per claim at $x\simeq0.80$ falls from $0.931$ to $0.216$. At $x\simeq1.20$, the number of shares sold short per claim falls from $0.843$ to $0.247$. Appendix~\ref{app:fin-grid} describes the discretization of the lognormal marginals. Table~\ref{tab:fin-bs} reports the seller's upper bound and the expected number of shares traded per claim for three of these values ($\eta=0.01,0.1,1$), together with the MOT and OT limiting prices.

\subsubsection{Prices and trading volumes with SPX marginals}
\label{subsec:fin-empirical}\label{subsec:fin-prices}

For the same forward-start payoff $\Phi_k$, we use Cboe quotes dated 9 October 2026 for AM-settled SPX calls and puts expiring on 20 November and 18 December. For each maturity $T_i$, we estimate the forward price $F_i$ and the discount factor $D_i$ from put--call parity. We divide quoted strikes by $F_i$ and option prices by $D_iF_i$. In these units, we fit the laws $\mu_i$ of $X_i=S_{T_i}/F_i$ jointly, requiring each $X_i$ to have expectation one and the fitted December call price to be at least the November price at each normalized strike. Appendix~\ref{subsec:fin-calibration} describes how we select the quotes and fit the two laws, and reports the pricing errors at each maturity.

\begin{figure}[H]
 \centering
 \includegraphics[width=\textwidth]{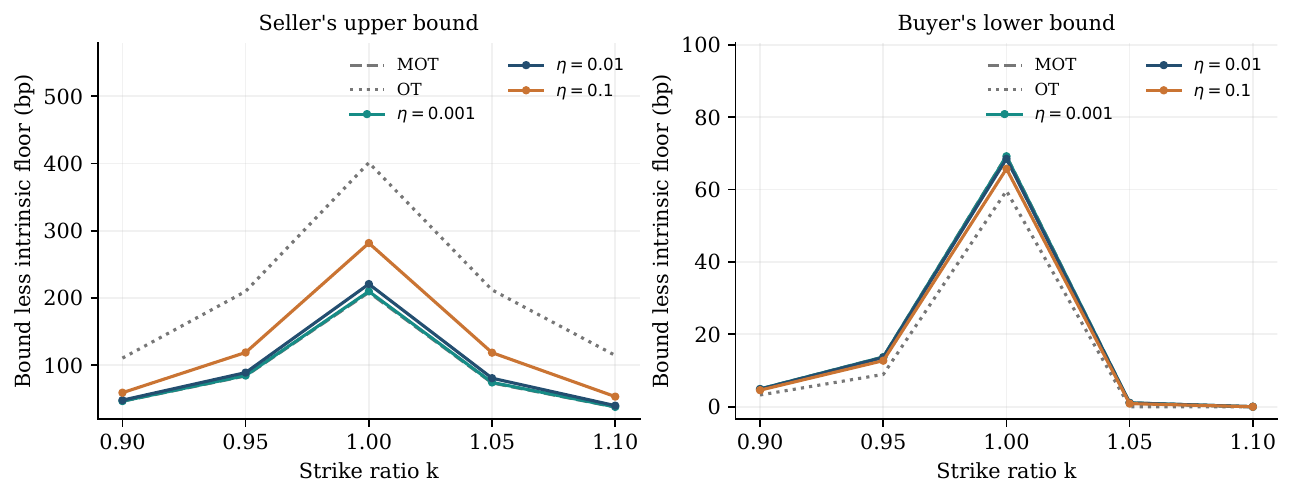}
\caption{Price bounds per claim for the forward-start call with marginals fitted to SPX option quotes. The left panel shows the seller's upper bound and the right panel the buyer's lower bound. To compare the bounds across strike ratios $k$, we subtract the common lower bound $(1-k)^+$ implied by $\E[X_1]=\E[X_2]=1$ and express the difference in normalized basis points. The solid curves labelled $\eta=0.001,0.01,0.1$ correspond to orders of $0.1$, $1$, and $10$ units, respectively. In each panel the dashed curve shows the MOT bound and the dotted curve the OT bound.}
 \label{fig:fin-empirical}
\end{figure}

We use these marginals for every order size and strike ratio and set $\eta_0=\kappa_1+\kappa_2=0.01$ for one claim. For an order of $n>0$ units of $\Phi_k$, the scaling relation \eqref{eq:asy-scale} implies that the seller's upper bound $U_n(k)$ and the buyer's lower bound $L_n(k)$ per claim satisfy
\begin{equation}\label{eq:fin-quantity}
 U_n(k)=\frac{P_\kappa(n\Phi_k)}{n}=P_{n\kappa}(\Phi_k),\qquad
 L_n(k)=-\frac{P_\kappa(-n\Phi_k)}{n}=-P_{n\kappa}(-\Phi_k).
\end{equation}
The effective impact per claim is therefore $\eta=n\eta_0$. Figure~\ref{fig:fin-empirical} compares the bounds for orders of $0.1$, $1$, and $10$ units at five strike ratios.

By Proposition~\ref{prop:asy-calculus}\textup{(iii)} and Theorem~\ref{thm:asy-endpoints}\textup{(ii)}, the price interval per claim approaches the MOT interval as the order size tends to zero and can only widen as the order grows. For $k=1$, increasing
the order from one to ten claims raises the seller's upper bound by about $61$ normalized basis points and reduces the buyer's lower bound by $2.76$. Under a law attaining the seller's bound, the expected number of shares traded rises for the whole order but falls per claim. Appendix~\ref{subsec:fin-calibration} reports the price bounds and trading volumes at this strike.

\subsection{Two-asset basket option}\label{subsec:fin-basket}

The forward-start call depends on one asset at two dates. We now consider a call on the sum of two asset prices at maturity and ask how the seller's upper bound changes when we interchange their impact coefficients. Calls on the individual assets fix four marginal laws. We optimize over joint laws with these marginals, since the expected basket payoff depends on how the terminal prices are paired.

We take $X_0=(0,0)$ and use the intervals in \cite[Section~4.1, footnote~4]{EGLO20}. For the first asset, the interval is $[-1,1]$ at $T_1$ and $[-3,3]$ at $T_2$. For the second asset, it is $[-3,3]$ at both dates. To form each marginal law, we divide its interval into 16 equal parts and place mass $1/16$ at each midpoint. The basket call pays $\Phi(X_1, X_2) = (X_2^1 + X_2^2)^+$.

On either asset, a call with strike zero and maturity $T_2$ costs $0.75$. Holding one call on each asset covers the basket payoff. Pairing equal terminal prices attains their combined price of $1.5$ in OT. Equality requires the two terminal prices to agree in sign. Under a martingale law, $X_2^2=X_1^2$ almost surely because the second asset has the same second moment at both dates. Equality would therefore make the sign of $X_2^1$ known at $T_1$. By the martingale property, the first asset's calls with strike zero would then have the same price for both maturities. These calls cost $0.25$ for maturity $T_1$ and $0.75$ for maturity $T_2$. The MOT upper bound for these discrete marginals is approximately $1.3764$.

\begin{figure}[H]
    \centering
    \includegraphics[width=\textwidth]{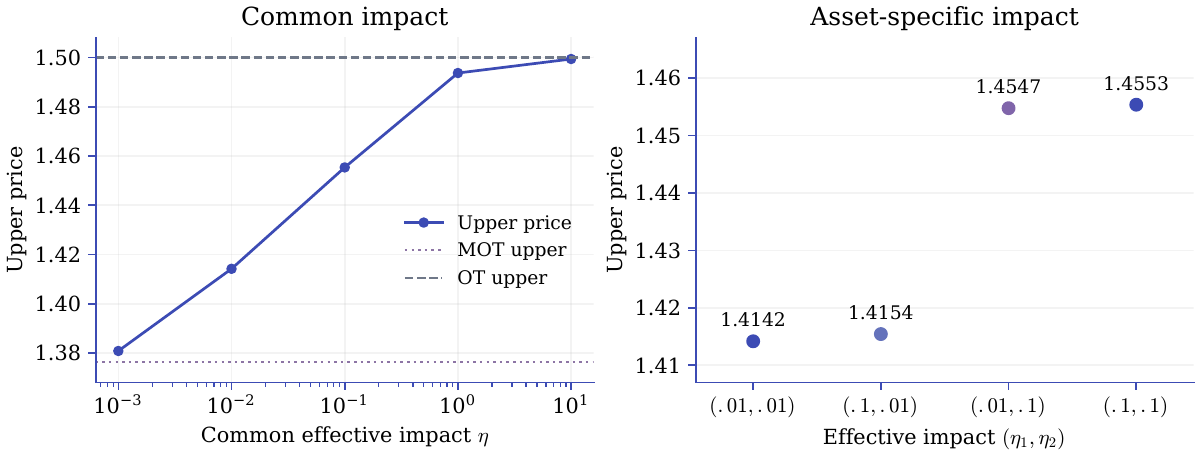}  \caption{Seller's upper bound for the basket call with 16 support points per marginal, in payoff units. The effective impact on asset $a$ is $\eta_a=\varepsilon(\kappa_1^a+\kappa_2^a)$. The left panel varies the common coefficient $\eta_1=\eta_2$ and marks the MOT and OT upper bounds. The right panel shows the upper bound for four pairs of impact coefficients. The two middle points correspond to $(\eta_1,\eta_2)=(0.1,0.01)$ and $(0.01,0.1)$, both with sum $0.11$.}
    \label{fig:fin-basket-prices}
\end{figure}

At $\eta_1=\eta_2=1$, the upper bound in Figure~\ref{fig:fin-basket-prices} is $1.4937$, close to its OT limit of $1.5$. With the sum of the coefficients fixed at $0.11$, assigning the larger coefficient to the second asset raises the bound from $1.4154$ to $1.4547$. This puts less weight on squared conditional drift in the asset whose price must be the same at both dates under MOT. Under the joint laws computed for these bounds the expected number of shares traded in that asset falls from $3.73$ to $1.07$ per claim, including liquidation. Appendix~\ref{app:basket-implementation} describes how we solve the discrete pricing problem and compute trading volumes from the resulting joint laws.

\section{Proofs}\label{sec:proof}
\subsection{Proofs for Section~\ref{sec:main}}

We first characterize the impact penalty under a fixed law and
show that its defining supremum can be taken over $\Hh_b$.
We then combine static superhedging duality with a minimax argument
to prove part~\textup{(i)} of Theorem~\ref{thm:main}.
The fixed-law characterization also yields part~\textup{(ii)}.

\subsubsection{The penalty under a fixed law}

Fix $\Q\in\mathcal P_1(\Omega)$. In \eqref{eq:penalty} we maximize expected trading gains net of impact costs over admissible strategies.
Write
$$b_i:=\E_\Q[X_{i+1}-X_i\mid\Fc_i],\qquad 0\leq i<N,$$
for the expected price increment given the history of all asset prices through date $i$. For $H\in\Hh_b$, we condition each price increment in \eqref{eq:gain} on $\Fc_i$ to obtain 
\begin{equation}\label{eq:fixed-objective}
  \E_\Q[G_\kappa(H)]
  =\sum_{i=0}^{N-1}\E_\Q[\langle b_i,H_i\rangle]
   -\frac12\E_\Q\sum_{i=0}^N
        \langle H_i - H_{i-1},K_i(H_i-H_{i-1})\rangle.
\end{equation}

Let $\mathcal U_\Q:=\prod_{i=0}^{N-1}L^2(\Fc_i,\Q;\R^d)$ be the space of square-integrable adapted holdings. For $h\in\mathcal U_\Q$, set $h_{-1}=h_N=0$. Since each difference $h_i-h_{i-1}$ is square integrable, the impact cost has finite expectation. If $b_i\in L^2(\Q;\R^d)$ for every $i<N$, the drift term is finite as well. The following proposition shows that in this case the supremum of the right-hand side of \eqref{eq:fixed-objective} over $h\in\mathcal U_\Q$ is attained and equals $I_\kappa(\Q)$.

\begin{proposition}\label{prop:fixed}
For every $\Q\in\mathcal P_1(\drev{\Omega})$,
\begin{equation}\label{eq:fixed-restriction}
 I_\kappa(\Q)
 =
 \sup_{H\in\Hh_b}\E_\Q[G_\kappa(H)].
\end{equation}
The value $I_\kappa(\Q)$ is finite if and only if
$b_i\in \drev{L^2(\Q;\R^d)}$ for every $i<N$.
In this case,
\begin{equation}\label{eq:fixed-max}
 I_\kappa(\Q)
 =
 \max_{h\in\mathcal U_\Q}
 \left\{
   \sum_{i=0}^{N-1}\E_\Q[\drev{\langle b_i,h_i\rangle}]
   -\frac12\E_\Q\sum_{i=0}^N
       \drev{\langle h_i-h_{i-1},K_i(h_i-h_{i-1})\rangle}
 \right\}.
\end{equation}
The maximizer $h^*\in\mathcal U_\Q$ is unique.
The process
$$M_i^*:=X_i+\drev{K_i(h_i^*-h_{i-1}^*)}, \qquad 0\leq i\leq N,$$
is an integrable $(\Fc_i)$-martingale under $\Q$, and
$$I_\kappa(\Q) = \frac12\E_\Q\sum_{i=0}^N \drev{\langle h_i^*-h_{i-1}^*,K_i(h_i^*-h_{i-1}^*)\rangle} = \frac12\E_\Q\sum_{i=0}^N \drev{\langle M_i^*-X_i,K_i^{-1}(M_i^*-X_i)\rangle}.$$
\end{proposition}

\begin{proof}
Set $$s:=\sup_{H\in\Hh_b}\E_\Q[G_\kappa(H)] \leq I_\kappa(\Q).$$
For each $i<N$, the space $C_b(\drev{\R^{d(i+1)};\R^d})$ is dense in $L^2\!\left(\drev{\R^{d(i+1)}},\,\Q\circ(X_0,\ldots,X_i)^{-1}\drev{;\R^d}\right).$ \borelrev{This follows from regularity of Borel probability measures on the Polish space $\drev{\R^{d(i+1)}}$.}
Since $\Fc_i=\sigma(X_0,\ldots,X_i)$, the image of $\Hh_b$ is therefore dense in $\mathcal U_\Q$ for the product $L^2$ norm. Write
$$
 \|h\|_{\mathcal U_\Q}^2
 :=\sum_{i=0}^{N-1}\E_\Q[\drev{|h_i|^2}],
 \qquad
 \|h\|_\kappa^2
 :=\E_\Q\sum_{i=0}^N\drev{\langle h_i-h_{i-1},K_i(h_i-h_{i-1})\rangle}.
$$
Cauchy--Schwarz applied to
$h_i=\sum_{j=0}^i(h_j-h_{j-1})$ gives
$$\frac{\drev{\min_{i,a}\kappa_i^a}}{N^2} \|h\|_{\mathcal U_\Q}^2 \leq  \|h\|_\kappa^2
 \leq 4\drev{\max_{i,a}\kappa_i^a}\, \|h\|_{\mathcal U_\Q}^2.$$
Thus $\|\cdot\|_\kappa$ is an equivalent Hilbert norm on $\mathcal U_\Q$.

Suppose first that $s<\infty$. Applying \eqref{eq:fixed-objective} to $tH$, with $H\in\Hh_b$ and $t\in\R$, and maximizing over $t$, we obtain
$$
 \left|\sum_{i=0}^{N-1}\E_\Q[\drev{\langle b_i,H_i\rangle}]\right|^2
 \leq 2s\|H\|_\kappa^2
 \leq
 8s\drev{\max_{i,a}\kappa_i^a}\,
       \|H\|_{\mathcal U_\Q}^2.
$$
Hence $H\mapsto\sum_{i=0}^{N-1}\E_\Q[\drev{\langle b_i,H_i\rangle}]$ extends to a continuous linear functional on $\mathcal U_\Q$. By the Riesz representation theorem, there exists $\beta=(\beta_0,\ldots,\beta_{N-1})\in\mathcal U_\Q$ such that
$$\sum_{i=0}^{N-1}\E_\Q[\drev{\langle b_i,H_i\rangle}] = \sum_{i=0}^{N-1}\E_\Q[\drev{\langle\beta_i,H_i\rangle}],\qquad H\in\Hh_b.$$
In the preceding identity, take $H$ zero except for one asset component at date $i$, which we set equal to $f(X_0, \ldots, X_i)$. Doing this for each component, we obtain
$$\E_\Q[b_i f(X_0,\ldots,X_i)] = \E_\Q[\beta_i f(X_0,\ldots,X_i)], \qquad f\in C_b(\drev{\R^{d(i+1)}}).$$
Since $b_i-\beta_i$ is integrable, the formula
$$
\rho_i(B):= \E_\Q\!\left[(b_i-\beta_i)\mathbf{1}_{\{(X_0,\ldots,X_i)\in B\}}\right], \qquad B\in\mathcal B(\drev{\R^{d(i+1)}}),
$$
defines a finite $\R^d$-valued Borel measure on $\R^{d(i+1)}$. By the preceding identity, $\int f\,d\rho_i=0$ for every $f\in C_b(\R^{d(i+1)})$, so $\rho_i=0$ by uniqueness of finite signed Borel measures applied to each component. Since $b_i-\beta_i$ is measurable with respect to $\Fc_i=\sigma(X_0,\ldots,X_i)$, it follows that $b_i=\beta_i$ $\Q$-almost surely. Thus $b_i\in L^2(\Q;\R^d)$ for every $i<N$.

Suppose now that $b_i\in \drev{L^2(\Q;\R^d)}$ for every $i<N$. The functional $v\mapsto\sum_{i=0}^{N-1}\E_\Q[\drev{\langle b_i,v_i\rangle}]$ is continuous for $\|\cdot\|_\kappa$. The Riesz representation theorem yields a unique $h^*\in\mathcal U_\Q$ satisfying
\begin{equation}\label{eq:fixed-first-order}
 \E_\Q\sum_{i=0}^N
   \drev{\langle K_i(h_i^*-h_{i-1}^*),v_i-v_{i-1}\rangle}
 =
 \sum_{i=0}^{N-1}\E_\Q[\drev{\langle b_i,v_i\rangle}],
 \qquad v\in\mathcal U_\Q.
\end{equation}
It follows that
$$\sum_{i=0}^{N-1}\E_\Q[\drev{\langle b_i,h_i\rangle}] - \frac12\|h\|_\kappa^2 = \frac12\|h^*\|_\kappa^2 - \frac12\|h-h^*\|_\kappa^2, \qquad h\in\mathcal U_\Q.$$
Thus $h^*$ uniquely maximizes this quadratic functional. By \eqref{eq:fixed-objective} and the preceding identity,
$$s = \frac12\|h^*\|_\kappa^2 -\frac12\inf_{H\in\Hh_b}\|H-h^*\|_\kappa^2 =\frac12\|h^*\|_\kappa^2,$$
where the last equality follows from the density of $\Hh_b$ in $\mathcal U_\Q$.

The process $M^*$ is adapted and integrable, since $\Q\in\mathcal P_1(\drev{\Omega})$ and $h_i^*-h_{i-1}^*\in \drev{L^2(\Q;\R^d)}\subset \drev{L^1(\Q;\R^d)}$ for $0\leq i\leq N$, with $h_{-1}^*=h_N^*=0$. Fix $i<N$ and let $\xi$ be a bounded $\Fc_i$-measurable \drev{$\R^d$-valued random vector}. Taking $v_i=\xi$ and $v_j=0$ for $j\ne i$ in
\eqref{eq:fixed-first-order} yields
$$\E_\Q\!\left[\drev{\xi\cdot}\left[\drev{K_i(h_i^*-h_{i-1}^*)} - \drev{K_{i+1}(h_{i+1}^*-h_i^*)} - b_i\right]\right]=0.$$
Since $\xi$ is arbitrary and $b_i$ and $h_i^*-h_{i-1}^*$ are $\Fc_i$-measurable, the defining property of conditional expectation gives
$$b_i = \drev{K_i(h_i^*-h_{i-1}^*)} - \drev{K_{i+1}} \E_\Q[h_{i+1}^*-h_i^*\mid\Fc_i], \qquad \Q\text{-a.s.}$$
Using $b_i=\E_\Q[X_{i+1}-X_i\mid\Fc_i]$, we obtain
$$\E_\Q[M_{i+1}^*-M_i^*\mid\Fc_i] = b_i + \drev{K_{i+1}} \E_\Q[h_{i+1}^*-h_i^*\mid\Fc_i] - \drev{K_i(h_i^*-h_{i-1}^*)}=0.$$
Thus $M^*$ is an integrable $(\Fc_i)$-martingale under $\Q$.

For every $H\in\Hh$ and every $\R^d$-valued adapted process $M=(M_i)_{i=0}^N$, we use $H_{-1}=H_N=0$, summation by parts in \eqref{eq:gain}, and Young's inequality to obtain
\begin{equation}\label{eq:gain-bound}
 G_\kappa(H)
 \leq\sum_{i=0}^{N-1}\drev{H_i\cdot(M_{i+1}-M_i)}
      +\frac12\sum_{i=0}^N\drev{\langle M_i-X_i,K_i^{-1}(M_i-X_i)\rangle},
 \qquad \Q\text{-a.s.}
\end{equation}
Fix $H\in\Hh$ and set
\begin{align*}
 A&:=\frac12\sum_{i=0}^N\drev{\langle M_i^*-X_i,K_i^{-1}(M_i^*-X_i)\rangle}
     =\frac12\sum_{i=0}^N\drev{\langle h_i^*-h_{i-1}^*,K_i(h_i^*-h_{i-1}^*)\rangle},\\
 T_n&:=\sum_{i=0}^{n-1}\drev{H_i\cdot(M_{i+1}^*-M_i^*)},
       \qquad 0\leq n\leq N.
\end{align*}
Then $A\geq0$, $A\in L^1(\Q)$ and $\E_\Q[A]=s$, while $T$
is a local martingale with $T_0=0$.
By \eqref{eq:gain-bound} with $M=M^*$,
$G_\kappa(H)\leq T_N+A$.
Admissibility and $\Q\in\mathcal P_1(\drev{\Omega})$ imply
\[
 T_N^-\leq A+G_\kappa(H)^-\in L^1(\Q).
\]
By \cite[Theorem~2]{JS98}, $T$ is a martingale, so $\E_\Q[T_N]=0$.
Taking expectations in the gain bound yields
$\E_\Q[G_\kappa(H)]\leq\E_\Q[A]=s$.
Taking the supremum over $H\in\Hh$ gives $I_\kappa(\Q)\leq s$.
Together with $s\leq I_\kappa(\Q)$, this proves
$I_\kappa(\Q)=s$.

\end{proof}

\subsubsection{Static superhedging duality}\label{subsec:static}

\begin{lemma}\label{lem:coupling-compactness}
Let $\drev{\mu_i^a\in\mathcal P_1(\R)}$ for $0\leq i\leq N$ \drev{and $1\leq a\leq d$}.
The set $\Pi(\mu)$ is nonempty, convex and weakly compact.
For every upper semicontinuous $f:\drev{\Omega}\to\R$ of linear growth,
the map $\Q\mapsto\E_\Q[f]$ is upper semicontinuous on
$\Pi(\mu)$, and is continuous when $f\in C_1(\drev{\Omega})$.
\end{lemma}

\begin{proof}
The product law $\bigotimes_{i,a}\mu_i^a$ belongs to $\Pi(\mu)$.
The marginal constraints are affine and are preserved under weak
convergence.

For each $(i,a)$, the law of $X_i^a$ is fixed and has finite first
moment. Thus the finite sum $\sum_{i,a}|X_i^a|$ is uniformly
integrable over $\Pi(\mu)$. Since $|X|\leq\sum_{i,a}|X_i^a|$, we
obtain, as in the proof of \cite[Lemma~2.2]{BHP13},
\begin{equation}\label{eq:tails}
  \lim_{R\to\infty}\sup_{\Q\in\Pi(\mu)}
  \int_{\{\max_{i,a}|x_i^a|>R\}}(1+|x|)\,\Q(dx)=0.
\end{equation}
For each $R>0$, the set $\{x\in\Omega:\max_{i,a}|x_i^a|\leq R\}$
is compact. Hence \eqref{eq:tails} implies that $\Pi(\mu)$ is tight.
Since $\Pi(\mu)$ is weakly closed, it is weakly compact by Prokhorov's theorem.

For upper semicontinuous $f$ of linear growth, denote the bounded truncations as $f_M:=(-M)\vee(f\wedge M)$. By \eqref{eq:tails}, $\sup_{\Q\in\Pi(\mu)}\E_\Q[|f-f_M|]\to0$ as $M\to\infty$.
For $\Q_n\Rightarrow\Q$ in $\Pi(\mu)$, applying the Portmanteau
theorem to $f_M$ and then letting $M\to\infty$ yields
$$\limsup_{n\to\infty}\E_{\Q_n}[f]\leq\E_\Q[f].$$
For $f\in C_1(\drev{\Omega})$, applying this inequality also to $-f$ gives $\E_{\Q_n}[f]\to\E_\Q[f]$.
\end{proof}

\begin{remark}\label{rem:weak-ui}
The expectation inequality and the convergence for $f\in C_1(\drev{\Omega})$ in Lemma~\ref{lem:coupling-compactness} can extend beyond the fixed-marginal setting. They remain valid for $\Q_n,\Q\in\mathcal P(\drev{\Omega})$ satisfying $\Q_n\Rightarrow\Q$, provided $1+|X|$ is uniformly integrable under $(\Q_n)_{n\geq1}$. Indeed, uniform integrability and Portmanteau imply that
$$\E_\Q[1+|X|] \leq \liminf_{n\to\infty}\E_{\Q_n}[1+|X|] \leq\sup_{n\geq1}\E_{\Q_n}[1+|X|]<\infty.$$
Thus $\Q\in\mathcal P_1(\drev{\Omega})$, and the truncation argument applies.
\end{remark}

For $f:\drev{\Omega}\to\R$ satisfying $|f(x)|\leq C(1+|x|)$ for some $C<\infty$, define its static superhedging cost by
\begin{equation}\label{eq:pi}
 \begin{aligned}
 \pi(f):=\inf\left\{
       \drev{\sum_{i=0}^N\sum_{a=1}^d\int_{\R}\varphi_i^a\,d\mu_i^a}:
       \varphi\in\Lambda, \,
       \drev{\sum_{i=0}^N\sum_{a=1}^d\varphi_i^a(x_i^a)}\geq f(x)
       \text{ for all }x\in \drev{\Omega}\right\}.
 \end{aligned}
\end{equation}
Let $\pi^{\mathcal V}(f)$ denote the value in \eqref{eq:pi} with static positions restricted to $\drev{\prod_{i=0}^N\prod_{a=1}^d\mathcal V_i^a}$.

\begin{proposition}\label{prop:static}
Assume that $\drev{\mu_i^a\in\mathcal P_1(\R)}$ for $0\leq i\leq N$ \drev{and $1\leq a\leq d$}. For every upper semicontinuous $f:\drev{\Omega}\to\R$ of at most linear growth,
\begin{equation}\label{eq:static-duality}
 \pi^{\mathcal V}(f)
 =\pi(f)
 =\max_{\Q\in\Pi(\mu)}\E_\Q[f]\in\R.
\end{equation}
\end{proposition}

\begin{proof}

For every $\Q\in\Pi(\mu)$ and every static superhedge $\varphi\in\Lambda$ of $f$,
$$\E_\Q[f] \leq \drev{\sum_{i=0}^N\sum_{a=1}^d\E_\Q[\varphi_i^a(X_i^a)]} = \drev{\sum_{i=0}^N\sum_{a=1}^d\int_{\R}\varphi_i^a\,d\mu_i^a}.$$
Weak duality and the inclusion $\drev{\mathcal V_i^a\subseteq C_1(\R)}$ therefore imply
\begin{equation}\label{eq:weak-dual-pi}
 \sup_{\Q\in\Pi(\mu)}\E_\Q[f]
 \leq\pi(f)\leq\pi^{\mathcal V}(f).
\end{equation}

The set $\Pi(\mu)$ consists of all couplings of the $d(N+1)$ one-dimensional marginals $\mu_i^a$. The dual functions in \cite[Proposition~2.1]{BHP13} are affine functions plus finite linear combinations of call payoffs. For each $(i,a)$, this class coincides with $\mathcal V_i^a$, since every affine payoff in $X_i^a$ is a linear combination of cash and the forward payoff $X_i^a-F_i^a$.

Apply that proposition to the lower semicontinuous cost $-f$ and write the dual functions as $-\varphi_i^a$, with $\varphi_i^a\in\mathcal V_i^a$. The pointwise constraint becomes $\sum_{i,a}\varphi_i^a(x_i^a)\geq f(x)$ for every $x\in\Omega$.
Hence
$$\pi^{\mathcal V}(f) = -\inf_{\Q\in\Pi(\mu)}\E_\Q[-f] = \sup_{\Q\in\Pi(\mu)}\E_\Q[f].$$
By \eqref{eq:weak-dual-pi}, $\pi(f) = \pi^{\mathcal V}(f)$. Lemma~\ref{lem:coupling-compactness} shows that the supremum over $\Pi(\mu)$ is finite and attained.
\end{proof}

\subsubsection{Minimax and the penalty representation}

We now combine the static duality with the fixed-law analysis.

\begin{proof}[Proof of Theorem~\ref{thm:main}]
For part~\textup{(i)}, write $D_\kappa^{\mathcal V}(\Phi)$ for the value in \eqref{eq:dual} with static positions restricted to $\drev{\prod_{i=0}^N\prod_{a=1}^d\mathcal V_i^a}$ and dynamic strategies restricted to $\Hh_b$. The linear growth of $\Phi$ and the finite first moments of the marginals $\mu_i^a$ imply $\sup_{\Q\in\Pi(\mu)}\E_\Q[|\Phi|]<\infty$.

Equip the vector space $\Hh_b$ with the norm
$\|H\|_\infty=\max_{0\leq i<N}\|h_i\|_\infty$.
By Lemma~\ref{lem:coupling-compactness}, $\Pi(\mu)$ is nonempty,
convex and weakly compact.
For each $H\in\Hh_b$, the residual $\Phi-G_\kappa(H)$ is upper
semicontinuous and has at most linear growth, since
$G_\kappa(H)\in C_1(\drev{\Omega})$.
Hence the map $\Q\mapsto\E_\Q[\Phi-G_\kappa(H)]$ is affine and
upper semicontinuous on $\Pi(\mu)$ by the same lemma.
For fixed $\Q\in\Pi(\mu)$, the map
$H\mapsto\E_\Q[\Phi-G_\kappa(H)]$ is convex, since
$H\mapsto G_\kappa(H)(x)$ is concave for every $x\in \drev{\Omega}$.
If $H^{(n)}\to H$ in $\|\cdot\|_\infty$, the strategies
$H^{(n)}$ are uniformly bounded. By \eqref{eq:gain}, their gains
converge pointwise to $G_\kappa(H)$ and admit a common linear
growth bound. Dominated convergence therefore yields continuity
of the objective in $H$.

Applying Proposition~\ref{prop:static} to $\Phi-G_\kappa(H)$ for each $H\in\Hh_b$, 
interchanging the infimum and supremum by Sion's minimax theorem, and identifying the penalty by \eqref{eq:fixed-restriction}, we obtain
\begin{align*}
 D_\kappa^{\mathcal V}(\Phi)
 &=\inf_{H\in\Hh_b}\max_{\Q\in\Pi(\mu)}
       \E_\Q[\Phi-G_\kappa(H)]\\
 &=\sup_{\Q\in\Pi(\mu)}\inf_{H\in\Hh_b}
       \E_\Q[\Phi-G_\kappa(H)]\\
 &=\sup_{\Q\in\Pi(\mu)}
       \bigl\{\E_\Q[\Phi]-I_\kappa(\Q)\bigr\}
   =P_\kappa(\Phi).
\end{align*}
The inner infimum is $-\infty$ when $I_\kappa(\Q)=\infty$,
since $\E_\Q[\Phi]\in\R$.
Weak duality and the inclusions of the restricted classes yield
\[
 P_\kappa(\Phi)\leq D_\kappa(\Phi)
 \leq D_\kappa^{\mathcal V}(\Phi)=P_\kappa(\Phi).
\]
Since $I_\kappa\geq0$, we have $P_\kappa(\Phi)<\infty$.
Moreover, Proposition~\ref{prop:fixed} implies that
$P_\kappa(\Phi)>-\infty$ if and only if some $\Q\in\Pi(\mu)$
satisfies \eqref{eq:finite-drift}.
The primal objective satisfies
\[
 \E_\Q[\Phi]-I_\kappa(\Q)
 =\inf_{H\in\Hh_b}\E_\Q[\Phi-G_\kappa(H)],
 \qquad \Q\in\Pi(\mu),
\]
and is therefore upper semicontinuous in $\Q$ as an infimum of the
upper semicontinuous maps considered above.
When $P_\kappa(\Phi)>-\infty$, this upper semicontinuous objective
attains its supremum on the weakly compact set $\Pi(\mu)$,
proving part~\textup{(i)}.

For part~\textup{(ii)}, fix $\Q\in\mathcal P_2(\drev{\Omega})$.
By conditional Jensen's inequality, \eqref{eq:finite-drift} holds.
Let $h^*$ and $M^*$ be as in Proposition~\ref{prop:fixed}.
The martingale $M^*$ belongs to $\mathcal M_2(\Q)$, since
$X_i$ and $h_i^*-h_{i-1}^*$ are square integrable for every $i$.
For $M\in\mathcal M_2(\Q)$ and $H\in\Hh_b$, the martingale
term in \eqref{eq:gain-bound} has zero expectation.
We take expectations in \eqref{eq:gain-bound} and then the supremum over $H\in\Hh_b$.
By \eqref{eq:fixed-restriction},
\[
 I_\kappa(\Q)
 \leq\frac12\E_\Q\sum_{i=0}^N\drev{\langle M_i-X_i,K_i^{-1}(M_i-X_i)\rangle},
 \qquad M\in\mathcal M_2(\Q).
\]
Equality holds at $M=M^*$ by the cost identity in
Proposition~\ref{prop:fixed}, establishing
\eqref{eq:penalty-representation}.
The minimizer is unique up to $\Q$-almost sure equality, since
the objective is strictly convex on the convex set
$\mathcal M_2(\Q)$.

Finally, for $\Q\in\mathcal P_1(\drev{\Omega})$, if $I_\kappa(\Q)=0$,
Proposition~\ref{prop:fixed} provides a martingale $M^*$ with
zero quadratic cost, hence $M_i^*=X_i$ $\Q$-almost surely
for every $i$, and $X$ is a $\Q$-martingale.
Conversely, if $X$ is a $\Q$-martingale, then $I_\kappa(\Q)=0$
by \eqref{eq:fixed-max}, since all conditional drifts vanish.
\end{proof}

\subsubsection{Dual attainment and cross-maturity no-arbitrage}
\label{subsec:attainment-na-proofs}

\begin{proof}[Proof of Proposition~\ref{prop:dual-attainment}]

Choose $R>0$ such that $A=[-R,R]$ contains the support of each $\mu_i^a$. Under the product coupling $\bigotimes_{i,a}\mu_i^a$, the price increments are bounded, so the conditional drifts satisfy \eqref{eq:finite-drift}. Hence, by Theorem~\ref{thm:main}\textup{(i)}, $D:=D_\kappa(\Phi)=P_\kappa(\Phi)\in\R$. We write $x_{0:i}=(x_0,\ldots,x_i)$ and keep the convention $H_{-1}=H_N=0$.

We restrict a minimizing sequence for \eqref{eq:dual} to $A^{d(N+1)}$ and truncate its holdings. The resulting bounds and the continuity of $\Phi$ allow us to choose hedges with equicontinuous static payoffs on $A$. After normalization, a subsequence of these payoffs converges uniformly to functions $\psi_i^a$ of total cost $D$. We then construct an adapted strategy for the residual claim $\Phi-\sum_{i,a}\psi_i^a(X_i^a)$ by backward recursion. To extend the resulting hedge to $\Omega$ at the same cost, we use the coordinate-wise projection $\tau(z):=(-R)\vee(z\wedge R)$ and add static payoffs that vanish on $A$.

\noindent\emph{Step 1.}
We truncate the holdings without decreasing the trading gain on $A^{d(N+1)}$. For a fixed asset $a$, write its total impact cost as
$$\frac12\sum_{i=0}^N\kappa_i^a(h_i-h_{i-1})^2 = \frac12 h^\top K^a h, \qquad h\in\R^N,$$
where $h_{-1}=h_N=0$ and $K^a$ is the symmetric $N\times N$ matrix defined by
$$(K^a h)_i = \kappa_i^a(h_i-h_{i-1}) + \kappa_{i+1}^a(h_i-h_{i+1}), \qquad 0\leq i<N.$$
The quadratic form is strictly positive for $h\ne0$, so $K^a$ is positive definite.

Since each price increment on $A^{d(N+1)}$ lies in $[-2R,2R]$, let $r^a$ solve $K^a r^a=2R\mathbf1$ and set $r_{-1}^a=r_N^a=0$. The holding cutoffs $r_i^a$ are positive for $0\leq i<N$. Indeed, if their minimum were nonpositive, choose an index $i$ attaining it. Since $r_{-1}^a=r_N^a=0$, we would have $r_i^a\leq r_{i-1}^a$ and $r_i^a \leq r_{i+1}^a$, and hence $(K^a r^a)_i\leq0$, contradicting $(K^a r^a)_i=2R>0$.

For $x\in A^{d(N+1)}$, write $h_i:=H_i^a(x)$ and $\bar h_i:=(-r_i^a)\vee(h_i\wedge r_i^a)$. If $h_i>r_i^a$, then $\bar h_i=r_i^a$, $\bar h_{i-1}\leq r_{i-1}^a$ and $\bar h_{i+1}\leq r_{i+1}^a$, so $(K^a\bar h)_i\geq2R\geq x_{i+1}^a-x_i^a$. If $h_i<-r_i^a$, the lower bounds imply $(K^a\bar h)_i\leq-2R\leq x_{i+1}^a-x_i^a$.
These inequalities and convexity of the impact cost yield
$$\sum_{i=0}^{N-1}(x_{i+1}^a-x_i^a)(h_i-\bar h_i) \leq\sum_{i=0}^{N-1} (K^a\bar h)_i(h_i-\bar h_i) \leq\frac12\bigl(h^\top K^a h -\bar h^\top K^a\bar h\bigr).$$

Rearranging this inequality shows that truncation does not decrease the net trading gain from asset $a$. Applying these cutoffs to each $H_i^a$ defines an adapted strategy $\bar H$. Summing over the assets gives $G_\kappa(\bar H)\geq G_\kappa(H)$ on $A^{d(N+1)}$.

Let $(\varphi,H)$ be a superhedge of $\Phi$ on $A^{d(N+1)}$, with $\varphi_i^a\in C(A)$ and $|H_i^a|\leq r_i^a$. Extend the strategy to $\Omega$ by
$$H_i^\tau(x_{0:i}) := H_i\bigl(\tau(x_0),\ldots,\tau(x_i)\bigr), \qquad x\in\Omega,\quad 0\leq i<N.$$
This extension is bounded and adapted, with a trading gain of at most linear growth. Thus $H^\tau\in\Hh$. Every $\Q\in\Pi(\mu)$ is supported on $A^{d(N+1)}$, where $H^\tau=H$. Taking expectations in the superhedging inequality therefore yields
$$\sum_{i,a}\int_A\varphi_i^a\,d\mu_i^a \geq \E_\Q[\Phi]-\E_\Q[G_\kappa(H^\tau)] \geq \E_\Q[\Phi]-I_\kappa(\Q).$$
Taking the supremum over $\Q\in\Pi(\mu)$ yields $\sum_{i,a}\int_A\varphi_i^a\,d\mu_i^a\geq P_\kappa(\Phi)=D$ by Theorem~\ref{thm:main}\textup{(i)}.

\noindent\emph{Step 2.} We regularize the static payoffs with moduli of continuity independent of the chosen hedge. Let $(\varphi,H)$ be a hedge of $\Phi$ on $A^{d(N+1)}$ with $\varphi_i^a\in C(A)$ and $|H_i^a|\leq r_i^a$. Then $|H_i^a-H_{i-1}^a|\leq r_{i-1}^a+r_i^a$ for $0\leq i\leq N$.

For $x\in A^{d(N+1)}$ and $s\in A$, let $x^{i,a,s}$ denote $x$ with replacing $x_i^a$ with $s$.
For $s,t\in A$, define
$$\rho_i^a(s,t) := \max_{x\in A^{d(N+1)}}|\Phi(x^{i,a,s})-\Phi(x^{i,a,t})|+(r_{i-1}^a+r_i^a)|s-t|.$$
With the holding values fixed, changing $x_i^a$ from $s$ to $t$ changes the linear trading gain by $(H_{i-1}^a-H_i^a)(t-s)$, whose absolute value is bounded by $(r_{i-1}^a+r_i^a)|s-t|$. Each $\rho_i^a$ is continuous and symmetric, vanishes on the diagonal and satisfies the triangle inequality.
Set
$$\psi_i^a(s) := \min_{t\in A}\{\varphi_i^a(t)+\rho_i^a(s,t)\},\qquad s\in A.$$

We use measurable selections to preserve feasibility under regularization, as in \cite[proof of Corollary~2.11]{BPRS25}. For each $s\in A$, let $T_i^a(s)$ be the smallest minimizer of $t\mapsto\varphi_i^a(t)+\rho_i^a(s,t)$ on $A$. Such a point exists by compactness and continuity. For $s_n\to s$ in $A$, compactness and joint continuity imply that every limit point of $T_i^a(s_n)$ is a minimizer at $s$. Hence $\liminf_{n\to\infty}T_i^a(s_n)\geq T_i^a(s)$, so $T_i^a$ is lower semicontinuous and therefore Borel measurable.
By construction,
$$\psi_i^a(s) = \varphi_i^a\bigl(T_i^a(s)\bigr) + \rho_i^a\bigl(s,T_i^a(s)\bigr), \qquad s\in A.$$

For $x\in A^{d(N+1)}$, define the path $y$ by $y_i^a:=T_i^a(x_i^a)$ and set
$$\widetilde H_i(x_{0:i}):=H_i(y_{0:i}), \qquad 0\leq i<N.$$
Since the maps $T_i^a$ are Borel and $y_{0:i}$ depends only on $x_{0:i}$, the strategy $\widetilde H$ is Borel adapted. It also satisfies $|\widetilde H_i^a|\leq r_i^a$. The holdings of $\widetilde H$ along $x$ equal those of $H$ along $y$, so their impact costs agree. Summation by parts gives
$$G_\kappa(\widetilde H)(x)-G_\kappa(H)(y) \geq-\sum_{i=0}^N\sum_{a=1}^d (r_i^a+r_{i-1}^a)|x_i^a-y_i^a|.$$
Compare $\Phi(x)$ and $\Phi(y)$ by changing one price coordinate at a time. Combining the definition of $\rho_i^a$ with the gain estimate and the superhedging inequality at $y$ yields
\begin{align*}
  \Phi(x)-G_\kappa(\widetilde H)(x)
  &\leq \Phi(y)-G_\kappa(H)(y)
       +\sum_{i,a}\rho_i^a(x_i^a,y_i^a)\\
  &\leq\sum_{i,a}
       \bigl[\varphi_i^a(y_i^a)+\rho_i^a(x_i^a,y_i^a)\bigr] =\sum_{i,a}\psi_i^a(x_i^a).
\end{align*}

The equality follows from the choice of $T_i^a$. Thus $(\psi,\widetilde H)$ is a superhedge on $A^{d(N+1)}$. Since $\psi_i^a\leq\varphi_i^a$, its static cost satisfies
$$\sum_{i,a}\int_A\psi_i^a\,d\mu_i^a\leq\sum_{i,a}\int_A\varphi_i^a\,d\mu_i^a.$$

Take a minimizing sequence $(\varphi^{(n)},H^{(n)})$ for \eqref{eq:dual}. Restrict these superhedges to $A^{d(N+1)}$ and apply the truncation from Step~1 followed by the regularization above. The resulting superhedges $(\psi^{(n)},\widetilde H^{(n)})$ satisfy $|\widetilde H_i^{a,(n)}|\leq r_i^a$ and
$$|\psi_i^{a,(n)}(s)-\psi_i^{a,(n)}(t)| \leq\rho_i^a(s,t), \qquad s,t\in A,\quad n\geq1.$$
Let $m_n:=\sum_{i,a}\int_A\psi_i^{a,(n)}\,d\mu_i^a$. By the lower bound from Step~1 and the preceding cost comparison,
$$D\leq m_n \leq\sum_{i,a}\int_A\varphi_i^{a,(n)}\,d\mu_i^a \longrightarrow D.$$

To obtain uniform bounds on the static payoffs, we normalize their means. Subtract $\int_A\psi_i^{a,(n)}\,d\mu_i^a$ from each $\psi_i^{a,(n)}$ and add $m_n$ to $\psi_0^{1,(n)}$, choosing $(0,1)$ as the reference index. The subtracted constants sum to $m_n$, so this operation preserves the total static payoff and its cost. Keeping the notation $\psi_i^{a,(n)}$ for the normalized functions, we have
$$\int_A\psi_i^{a,(n)}\,d\mu_i^a =
  \begin{cases}
    m_n, & (i,a)=(0,1),\\
    0,   & (i,a)\ne(0,1).
  \end{cases}
$$
Adding constants preserves the modulus estimate. Hence, for every $s\in A$,
$$|\psi_i^{a,(n)}(s)| \leq \left|\int_A\psi_i^{a,(n)}\,d\mu_i^a\right| + \int_A\rho_i^a(s,t)\,\mu_i^a(dt) \leq |m_n|+\max_{u,v\in A}\rho_i^a(u,v).$$
Since $m_n\to D$, each sequence $(\psi_i^{a,(n)})_{n\geq1}$ is uniformly bounded and equicontinuous on $A$.

Since there are only finitely many pairs $(i,a)$, the Arzel\`a--Ascoli
theorem yields a common subsequence, still indexed by $n$, and functions $\psi_i^a\in C(A)$ such that
$$\psi_i^{a,(n)}\longrightarrow\psi_i^a \quad\text{uniformly on }A \quad\text{for every }(i,a).$$
By uniform convergence,
$$\sum_{i,a}\int_A\psi_i^a\,d\mu_i^a = \lim_{n\to\infty}m_n=D.$$

\noindent\emph{Step 3.}
We construct a dynamic strategy for the limiting static payoffs on $A^{d(N+1)}$. Put
$$\mathcal B_i:=\prod_{a=1}^d[-r_i^a,r_i^a], \qquad 0\leq i<N.$$
For the limiting static portfolio $\psi$, the value $V_i(x_{0:i},h)$ represents the capital required to cover the remaining payoff from date $i$ onward, given the price history $x_{0:i}$ and the holding $h\in\R^d$ before the date-$i$ trade. Starting with terminal liquidation, define
$$V_N(x_{0:N},h) := \Phi(x)-\sum_{i,a}\psi_i^a(x_i^a) +\frac12\langle h,K_Nh\rangle,$$
and, for $i=N-1,\ldots,0$,
$$V_i(x_{0:i},h) :=\min_{v\in\mathcal B_i}\max_{z\in A^d} \Bigl\{\frac12\langle v-h,K_i(v-h)\rangle -v\cdot(z-x_i) +V_{i+1}(x_{0:i},z,v)\Bigr\}.$$
For each $n$, define $V_i^{(n)}$ by the same recursion, replacing
$\psi_i^a$ in the terminal condition with $\psi_i^{a,(n)}$.

Since the terminal conditions are continuous and both optimization sets are fixed and compact, backward induction shows that $V_i$ and $V_i^{(n)}$ are continuous. At each date, a minimizing holding for $V_i$ can be selected as a Borel function of $(x_{0:i},h)$, since the objective is continuous and $\mathcal B_i$ is fixed and compact.

Substituting the holdings $\widetilde H_i^{(n)}\in\mathcal B_i$ from Step~2 into the successive minima gives, for every $x_0\in A^d$,
$$V_0^{(n)}(x_0,0) \leq \sup_{(x_1,\ldots,x_N)\in A^{dN}} \Bigl\{\Phi(x)-\sum_{i,a}\psi_i^{a,(n)}(x_i^a) - G_\kappa(\widetilde H^{(n)})(x) \Bigr\} \leq0, $$
where the last inequality follows from the superhedging property of $(\psi^{(n)},\widetilde H^{(n)})$ on $A^{d(N+1)}$. Taking maxima and minima does not enlarge a uniform difference. Since the terminal conditions differ only through the static payoffs, we obtain
$$
  \bigl\|V_0^{(n)}(\cdot,0)-V_0(\cdot,0)\bigr\|_{\infty,A^d}
  \leq
  \sum_{i,a}
  \bigl\|\psi_i^{a,(n)}-\psi_i^a\bigr\|_{\infty,A}
  \longrightarrow0
$$
by the uniform convergence in Step~2. Consequently, $V_0(x_0,0)\leq0$ for every $x_0\in A^d$.

Set $H_{-1}^A=H_N^A=0$. For $i=0,\ldots,N-1$, evaluate a Borel minimizing selection for $V_i$ at $(x_{0:i},H_{i-1}^A)$ to define $H_i^A(x_{0:i})$. The resulting strategy is Borel adapted and satisfies $H_i^A\in\mathcal B_i$. Summing the recursive inequalities along this strategy yields
$$\Phi(x)-\sum_{i,a}\psi_i^a(x_i^a)-G_\kappa(H^A)(x) \leq V_0(x_0,0)\leq0, \qquad x\in A^{d(N+1)}.$$

\noindent\emph{Step 4.}
We extend the hedge to $\Omega$ using static corrections that vanish on $A$. Set $H_i^*(x_{0:i}):=H_i^A(\tau(x_{0:i}))$ for $0\leq i<N$. The extension argument in Step~1 shows $H^*\in\Hh$. Comparing the gains along $x$ and $\tau x$ by summation by parts, as in Step~2, yields
$$G_\kappa(H^*)(x)-G_\kappa(H^A)(\tau x) \geq -\sum_{i=0}^N\sum_{a=1}^d (r_i^a+r_{i-1}^a)|x_i^a-\tau(x_i^a)|, \qquad x\in\Omega.$$

To control the change in the claim payoff, define
$$\omega(t):= \max_{z\in[-R-t,R+t]^{d(N+1)}} \bigl(\Phi(z)-\Phi(\tau z)\bigr)^+, \qquad t\geq0.$$
Rescaling the maximization to the fixed cube $[-1,1]^{d(N+1)}$ shows that $\omega$ is continuous. It is nonnegative and nondecreasing, with $\omega(0)=0$, and has at most linear growth by the growth assumption on $\Phi$. For every $x\in\Omega$,
$$\Phi(x)-\Phi(\tau x) \leq \omega\!\left(\max_{i,a}|x_i^a-\tau(x_i^a)|\right) \leq
  \sum_{i,a}\omega\bigl(|x_i^a-\tau(x_i^a)|\bigr).$$

Set, for $s\in\R$, $\varphi_i^{a,*}(s) := \psi_i^a(\tau(s)) + \omega\bigl(|s-\tau(s)|\bigr) +(r_i^a+r_{i-1}^a)|s-\tau(s)|.$ These functions are continuous and have at most linear growth, so $\varphi^*\in\Lambda$. Combining the hedge inequality from Step~3 at $\tau x$ with the two estimates above, we obtain
$$\sum_{i,a}\varphi_i^{a,*}(x_i^a) + G_\kappa(H^*)(x)\geq\Phi(x), \qquad x\in\Omega.$$
Since $\varphi_i^{a,*}=\psi_i^a$ on $A$ and $\mu_i^a(A)=1$, their total cost is
$$\sum_{i,a}\int_{\R}\varphi_i^{a,*}\,d\mu_i^a = \sum_{i,a}\int_A\psi_i^a\,d\mu_i^a =D.$$
\end{proof}

\begin{proof}[Proof of Proposition~\ref{prop:convex-order-na}]
If $\mathrm{NA}_\kappa$ holds, then $P_\kappa(0)=D_\kappa(0)=0$. By Theorem~\ref{thm:main}\textup{(i)}, there exists some $\Q \in \Pi(\mu)$ attains this finite value and therefore satisfies $I_\kappa(\Q) = 0$. Conversely, if $I_\kappa(\Q) = 0$ for some $\Q \in \Pi(\mu)$, then $D_\kappa(0)=0$ since $I_\kappa\geq0$. By Theorem~\ref{thm:main}\textup{(ii)}, for $\Q \in \Pi(\mu)$, we have $I_\kappa(\Q) = 0$ exactly when $X$ is a $\Q$-martingale.

Under any such law, conditional Jensen's inequality implies $\mu_i^a\cv\mu_{i+1}^a$ for every $i<N$ and every asset $a$. Conversely, suppose these convex orders hold. For each asset $a$, take a martingale coupling of each successive pair of marginals by \cite[Theorem~8]{Strassen1965}. Composing their transition kernels yields a scalar martingale path law with marginals $(\mu_i^a)_{i=0}^N$. Take the product of these path laws over the assets. Under the product law $\Q$, the other assets' histories are independent of the entire path of asset $a$. Consequently $\E_\Q[X_{i+1}^a\mid\Fc_i] =\E_\Q[X_{i+1}^a\mid X_0^a,\ldots,X_i^a]=X_i^a$ for every $i<N$ and $a$. Thus $\Q\in\mathcal M(\mu)$.
\end{proof}

\subsection{Proofs for Section~\ref{sec:one-period}}
\label{subsec:one-period-proofs}

\begin{proof}[Proof of Proposition~\ref{prop:one-period-formulation}]
Fix $\gamma\in\Pi(\mu_0,\mu_1)$. If $b_\gamma\in L^2(\mu_0)$, Proposition~\ref{prop:fixed} with $d=N=1$ reduces the penalty to a quadratic maximization over $L^2(\mu_0)$. Completing the square yields
$$I_\kappa(\gamma)= \max_{h\in L^2(\mu_0)}\int_{\R} \left(hb_\gamma-\frac{\kappa_0+\kappa_1}{2}h^2\right)d\mu_0 = \frac{1}{2(\kappa_0+\kappa_1)}\int_{\R} b_\gamma^2\,d\mu_0.$$
If $b_\gamma\notin L^2(\mu_0)$, Proposition~\ref{prop:fixed} implies $I_\kappa(\gamma)=+\infty$. This proves \eqref{eq:wot-barycentric} for every $\gamma \in \Pi(\mu_0, \mu_1)$.

The growth bound on $\Phi$ and the marginal first moments imply $\Phi\in L^1(\gamma)$. Disintegrating $\gamma$ therefore gives
$$\int_{\R}C_\kappa^\Phi(x,\gamma_x)\,\mu_0(dx) = I_\kappa(\gamma)-\E_\gamma[\Phi].$$
Equation~\eqref{eq:primal} and Theorem~\ref{thm:main}\textup{(i)} now imply \eqref{eq:one-period-cost}, the finiteness criterion and attainment when the value is finite.
\end{proof}

\begin{proof}[Proof of Theorem~\ref{thm:one-period-structure}]
\textup{(i)} By Proposition~\ref{prop:one-period-formulation}, the finite value is attained at some $\gamma^*$ with $b_\kappa^*:=b_{\gamma^*}\in L^2(\mu_0)$. For a coupling $\gamma$ with $b_\gamma\in L^2(\mu_0)$ and $0<t<1$, set $\gamma_t=(1-t)\gamma^*+t\gamma$. Since both couplings have first marginal $\mu_0$, their mixture has drift $b_{\gamma_t}=(1-t)b_\kappa^*+t b_\gamma$. Expanding its squared norm in \eqref{eq:wot-barycentric} and using linearity of the payoff integral, we obtain
\begin{align*}
 0&\le P_\kappa(\Phi)-\bigl(\E_{\gamma_t}[\Phi]-I_\kappa(\gamma_t)\bigr)\\
  &=t\bigl[P_\kappa(\Phi)-(\E_\gamma[\Phi]-I_\kappa(\gamma))\bigr]
    -\frac{t(1-t)}{2(\kappa_0+\kappa_1)}\|b_\gamma-b_\kappa^*\|_{L^2(\mu_0)}^2.
\end{align*}
Divide by $t$ and let $t\downarrow0$ to obtain \eqref{eq:one-period-gap}. For $b_\gamma\notin L^2(\mu_0)$ the right-hand side is $+\infty$. For any maximizing coupling it is zero, hence every maximizer has drift $b_\kappa^*$ $\mu_0$-a.s. This proves uniqueness in $L^2(\mu_0)$.

\textup{(ii)} For $\mu_0,\mu_1\in\mathcal P_2(\R)$, conditional Jensen's inequality yields $\|b_\gamma\|_{L^2(\mu_0)}^2\leq\E_\gamma|X_1-X_0|^2<\infty$ for every $\gamma \in \Pi(\mu_0, \mu_1)$. Thus the value is finite by Proposition~\ref{prop:one-period-formulation}, and \textup{(i)} applies.

If $\Gamma\in\mathcal A(\zeta)$, then $\E_\Gamma[X_1\mid Z]=Z$ by iterated conditioning, hence $\zeta\cv\mu_1$ by conditional Jensen's inequality. Conversely, suppose that $\zeta\cv\mu_1$. Then $\zeta$ has finite second moment by convexity of $z\mapsto z^2$, and Strassen's theorem \cite[Theorem~8]{Strassen1965} provides a martingale coupling $\zeta(dz)k_z(dy)$ with terminal marginal $\mu_1$. For every $\rho\in\Pi(\mu_0,\zeta)$, the law $\Gamma(dx,dz,dy)=\rho(dx,dz)k_z(dy)$ extends $\rho$ and satisfies $\E_\Gamma[X_1\mid X_0,Z]=Z$. All three marginals have finite second moments, hence $\Gamma\in\mathcal A(\zeta)$.

Fix $\Gamma\in\mathcal A(\zeta)$ and let $\gamma$ be its $(X_0,X_1)$-marginal. Conditioning $\E_\Gamma[X_1\mid X_0,Z]=Z$ on $X_0$ yields $\E_\Gamma[Z\mid X_0]=\E_\Gamma[X_1\mid X_0]=X_0+b_\gamma(X_0)$. Thus $Z-X_0-b_\gamma(X_0)$ is orthogonal to $(b_\gamma-h)(X_0)$ for every $h\in L^2(\mu_0)$, and
\begin{equation}\label{eq:one-period-orthogonal}
    \E_\Gamma|Z-X_0-h(X_0)|^2 = \|b_\gamma-h\|_{L^2(\mu_0)}^2+\E_\Gamma|Z-X_0-b_\gamma(X_0)|^2.
\end{equation}
Using \eqref{eq:one-period-orthogonal} first with $h=0$ and then with
$h=b_\kappa^*$, we obtain
\begin{align*}
 P_\kappa(\Phi)-\E_\Gamma\!\left[\Phi(X_0,X_1) - \frac{|X_0-Z|^2}{2(\kappa_0+\kappa_1)}\right]
 &=P_\kappa(\Phi)-\E_\gamma[\Phi]+I_\kappa(\gamma)
          +\frac{\E_\Gamma|Z-X_0-b_\gamma(X_0)|^2}{2(\kappa_0+\kappa_1)}\\
 &\geq\frac{\E_\Gamma|Z-X_0-b_\kappa^*(X_0)|^2}{2(\kappa_0+\kappa_1)}
       \geq\frac{\Wc_2(\zeta,\zeta^*)^2}{2(\kappa_0+\kappa_1)}.
\end{align*}
The equality follows from \eqref{eq:wot-barycentric}. The first inequality uses \eqref{eq:one-period-gap}, whose right-hand side equals $P_\kappa(\Phi)-\E_\gamma[\Phi]+I_\kappa(\gamma)$ by Proposition~\ref{prop:one-period-formulation}. The last uses the coupling $(Z,X_0+b_\kappa^*(X_0))$ of $\zeta$ and $\zeta^*=(\mathrm{id}+b_\kappa^*)_\#\mu_0$. By \eqref{eq:one-period-value}, the gap vanishes precisely when $\gamma$ maximizes the endpoint objective and $Z=X_0+b_\gamma(X_0)$ $\Gamma$-a.s.

For each maximizer $\gamma^*$ of \eqref{eq:one-period-value}, set $Z=X_0+b_\kappa^*(X_0)$ under $\gamma^*$ using a Borel version of the drift. By \textup{(i)}, $Z=\E_{\gamma^*}[X_1\mid X_0]$ is square integrable and $\E_{\gamma^*}[X_1\mid X_0,Z]=Z$ since $Z$ is a function of $X_0$. The law $\Gamma^*$ of $(X_0,Z,X_1)$ thus belongs to $\mathcal A(\zeta^*)$. By the equality criterion and the upper bound,
\begin{align*}
 P_\kappa(\Phi)
 &=\E_{\Gamma^*}\!\left[\Phi(X_0,X_1)-\frac{|X_0-Z|^2}{2(\kappa_0+\kappa_1)}\right]\\
 &\leq\sup_{\zeta\cv\mu_1}\ \sup_{\Gamma\in\mathcal A(\zeta)}
       \E_\Gamma\!\left[\Phi(X_0,X_1)-\frac{|X_0-Z|^2}{2(\kappa_0+\kappa_1)}\right]
 \leq P_\kappa(\Phi).
\end{align*}
This proves \eqref{eq:one-period-lift} with attainment at $(\zeta^*,\Gamma^*)$. For fixed $\zeta$, the inner supremum is at most $P_\kappa(\Phi)-\Wc_2(\zeta,\zeta^*)^2/[2(\kappa_0+\kappa_1)]$, hence $\zeta^*$ is the unique outer maximizer. At $\zeta^*$, any inner maximizer $\Gamma$ has an optimal endpoint marginal $\gamma$ and satisfies $Z=X_0+b_\kappa^*(X_0)$ $\Gamma$-a.s. by the equality criterion and \textup{(i)}. Thus the inner maximizers are exactly the laws constructed above.

\textup{(iii)} For $\Phi=0$, the objective in \eqref{eq:one-period-lift} depends only on the $(X_0,Z)$-marginal. Fix $\zeta\cv\mu_1$. By the extension property in \textup{(ii)}, this marginal ranges over all of $\Pi(\mu_0,\zeta)$ as $\Gamma$ ranges over $\mathcal A(\zeta)$. Using $P_\kappa(0)=D_\kappa(0)$ from Proposition~\ref{prop:one-period-formulation}, we obtain
\begin{align*}
 P_\kappa(0)=D_\kappa(0)
 &=-\frac{1}{2(\kappa_0+\kappa_1)}
    \inf_{\zeta\cv\mu_1}\inf_{\rho\in\Pi(\mu_0,\zeta)}
    \int_{\R^2}|x-z|^2\,\rho(dx,dz)\\
 &=-\frac{1}{2(\kappa_0+\kappa_1)}
    \inf_{\zeta\cv\mu_1}\Wc_2(\mu_0,\zeta)^2.
\end{align*}
The unique outer maximizer in \textup{(ii)} is therefore the unique minimizer $\mu_0^*=\zeta^*=(\mathrm{id}+b_\kappa^*)_\#\mu_0$ in \eqref{eq:one-period-projection}. The Wasserstein minimization depends only on $\mu_0$ and $\mu_1$, hence $\mu_0^*$ is the same for every $\kappa\in(0,\infty)^2$. If $\mu_0\cv\mu_1$, then $\zeta=\mu_0$ is feasible with zero cost, hence $\mu_0^*=\mu_0$ by uniqueness. The converse follows from $\mu_0^*\cv\mu_1$.

The maximizers of \eqref{eq:one-period-value} now minimize the quadratic barycentric cost $\int_{\R}b_\gamma^2\,d\mu_0$. Backhoff-Veraguas et al.~\cite[Theorem~1.4]{BBP19} show that there is a convex $\varphi\in C^1(\R)$ with $1$-Lipschitz derivative such that, for every such maximizer $\gamma^*$,
$$\varphi'(x)=\int_{\R}y\,\gamma_x^*(dy)=x+b_\kappa^*(x) \qquad\text{for }\mu_0\text{-a.e.\ }x,$$
where the last equality follows from \textup{(i)}. Since $\varphi$ is convex, $\varphi'$ is nondecreasing.

\end{proof}

\keeprevisioncolorstrue%
\subsection{Proofs for Section~\ref{subsec:asymptotics}}

\begin{proof}[Proof of Proposition~\ref{prop:asy-calculus}]
\textup{(i)} For every Borel adapted strategy $H$, the gain formula \eqref{eq:gain} implies
$$G_{\varepsilon\kappa}(H/\varepsilon) = \varepsilon^{-1}G_\kappa(H).$$
The lower bound in \eqref{eq:admissible} scales by the same factor. By \eqref{eq:admissible}, $H$ is admissible for $\kappa$ if and only if $H/\varepsilon$ is admissible for $\varepsilon\kappa$. Taking expectations and then suprema proves \eqref{eq:asy-scale}, including the case of infinite penalty.

\textup{(ii)} Suppose $b=(b_i)_{i<N}\in\mathcal U_\Q$. For
$h\in\mathcal U_\Q$ with $h_{-1}=h_N=0$, the norm bounds in the proof
of Proposition~\ref{prop:fixed} read
\begin{equation}\label{eq:asy-quadratic}
 \frac{\min_{j,a}\kappa_j^a}{N^2}\|h\|_{\mathcal U_\Q}^2
 \leq\|h\|_\kappa^2
 \leq4\max_{j,a}\kappa_j^a\,\|h\|_{\mathcal U_\Q}^2.
\end{equation}
Test \eqref{eq:fixed-max} at $h_i=b_i/(4\max_{j,a}\kappa_j^a)$. By \eqref{eq:asy-quadratic}, the cost $\|h\|_\kappa^2/2$ is at most half the drift term for this choice of $h$, proving the lower bound in \eqref{eq:asy-drift}. For the upper bound, Young's inequality and
\eqref{eq:asy-quadratic} imply
$$
 \sum_{i=0}^{N-1}\E_\Q[\langle b_i,h_i\rangle]
 \leq\frac{N^2}{2\min_{j,a}\kappa_j^a}
       \sum_{i=0}^{N-1}\E_\Q[|b_i|^2]+\frac12\|h\|_\kappa^2.
$$
Subtracting $\|h\|_\kappa^2/2$ and maximizing over $h$ proves the upper bound in \eqref{eq:asy-drift}.

\textup{(iii)} For $H\in\Hh_b$, the gain $G_\kappa(H)$ belongs to $C_1(\Omega)$. Hence $\Q\mapsto\E_\Q[G_\kappa(H)]$ is affine and weakly continuous on $\Pi(\mu)$ by Lemma~\ref{lem:coupling-compactness}. By \eqref{eq:fixed-restriction}, $I_\kappa$ is the supremum of these functionals and is therefore convex and weakly lower semicontinuous.

Substitute \eqref{eq:asy-scale} into \eqref{eq:primal} and apply Theorem~\ref{thm:main}\textup{(i)} at impact $\varepsilon\kappa$. Since $I_\kappa\geq0$, the objective at each law is nondecreasing in $\varepsilon$. Expected payoffs are bounded above on $\Pi(\mu)$ by the linear growth of $\Phi$ and the fixed first moments. A law of finite penalty also supplies a finite lower bound on the value, and Theorem~\ref{thm:main}\textup{(i)} then ensures attainment.
\end{proof}

\begin{proof}[Proof of Theorem~\ref{thm:asy-endpoints}]
The second moments of the marginals bound $I_\kappa$ uniformly over $\Pi(\mu)$. Indeed, for each $\Q\in\Pi(\mu)$, the process $M_i=X_0$ for $0\leq i\leq N$ is a square-integrable martingale. Using $M$ as a competitor in \eqref{eq:penalty-representation}, we obtain
\begin{equation}\label{eq:asy-uniform}
 0\leq I_\kappa(\Q)
 \leq\frac12\sum_{i=1}^N\sum_{a=1}^d
       \frac{\E_\Q[(X_i^a-X_0^a)^2]}{\kappa_i^a}
 \leq\sum_{i=1}^N\sum_{a=1}^d
       \frac{\int t^2\mu_i^a(dt)+\int t^2\mu_0^a(dt)}{\kappa_i^a}<\infty,
\end{equation}
Since $I_\kappa$ is finite on $\Pi(\mu)$, Proposition~\ref{prop:asy-calculus}\textup{(iii)} ensures that the supremum defining $P_{\varepsilon\kappa}(\Phi)$ is attained for every $\varepsilon>0$.

\textup{(i)} By Lemma~\ref{lem:coupling-compactness}, the laws attaining $O$ form a nonempty weakly compact set. The minimum defining $j_O$ in
Proposition~\ref{prop:rate-OT} is therefore attained by lower semicontinuity of $I_\kappa$ and is finite by \eqref{eq:asy-uniform}. At a minimizing law the penalized objective equals
$O-j_O/\varepsilon$. By optimality of $\Q_\varepsilon$,
\begin{equation}\label{eq:asy-OT-comparison}
 0\leq\varepsilon\bigl(O-P_{\varepsilon\kappa}(\Phi)\bigr)
 =\varepsilon\bigl(O-\E_{\Q_\varepsilon}[\Phi]\bigr)
       +I_\kappa(\Q_\varepsilon)
 \leq j_O.
\end{equation}
In particular, $P_{\varepsilon\kappa}(\Phi)\to O$ as $\varepsilon\to\infty$. The inequalities $P_{\varepsilon\kappa}(\Phi)\leq\E_{\Q_\varepsilon}[\Phi]\leq O$ then imply that $\E_{\Q_\varepsilon}[\Phi]\to O$. Along any sequence $\varepsilon_n\to\infty$ for which $\Q_{\varepsilon_n}\Rightarrow\Q_\infty$, upper semicontinuity of the payoff integral implies $\E_{\Q_\infty}[\Phi]=O$. Lower semicontinuity of $I_\kappa$ and \eqref{eq:asy-OT-comparison} now yield
$$
 j_O\leq I_\kappa(\Q_\infty)
 \leq\liminf_{n\to\infty}I_\kappa(\Q_{\varepsilon_n})
 \leq\limsup_{n\to\infty}I_\kappa(\Q_{\varepsilon_n})\leq j_O.
$$
Thus $\Q_\infty$ has the least penalty among the laws attaining $O$. For any $\varepsilon_n\to\infty$, weak compactness ensures that every subsequence of $(\Q_{\varepsilon_n})$ has a weakly convergent further subsequence. The penalties tend to $j_O$ along each such further subsequence, hence $I_\kappa(\Q_\varepsilon)\to j_O$ as $\varepsilon\to\infty$.

\textup{(ii)} Let
$$j_*:=\min_{\Q\in\Pi(\mu)}I_\kappa(\Q),\qquad m_*:=\max_{\substack{\Q\in\Pi(\mu)\\I_\kappa(\Q)=j_*}}\E_\Q[\Phi].$$
The minimum is finite and attained by \eqref{eq:asy-uniform}, compactness and lower semicontinuity. Its minimizers form a compact set on which the payoff integral attains $m_*$. A law attaining $m_*$ has penalized value $m_*-j_*/\varepsilon$, hence
\begin{equation}\label{eq:asy-small-comparison}
    m_* \leq P_{\varepsilon\kappa}(\Phi)+\frac{j_*}{\varepsilon} =\E_{\Q_\varepsilon}[\Phi]
       -\frac{I_\kappa(\Q_\varepsilon)-j_*}{\varepsilon}
    \leq\E_{\Q_\varepsilon}[\Phi]\leq O.
\end{equation}
In particular, $0\leq I_\kappa(\Q_\varepsilon)-j_*\leq\varepsilon(O-m_*)$. Take any sequence $\varepsilon_n\downarrow0$ along which $\Q_{\varepsilon_n}\Rightarrow\Q_0$. Lower semicontinuity implies $I_\kappa(\Q_0)=j_*$. By \eqref{eq:asy-small-comparison} and upper semicontinuity of the payoff integral,
$$
 m_*\leq\liminf_{n\to\infty}\E_{\Q_{\varepsilon_n}}[\Phi]
 \leq\limsup_{n\to\infty}\E_{\Q_{\varepsilon_n}}[\Phi]
 \leq\E_{\Q_0}[\Phi]\leq m_*.
$$
Thus every such $\Q_0$ maximizes expected payoff among the penalty minimizers, and compactness implies $\E_{\Q_\varepsilon}[\Phi]\to m_*$.

By Theorem~\ref{thm:main}\textup{(ii)}, $I_\kappa$ vanishes exactly on $\mathcal M(\mu)$. If this set is nonempty, $j_*=0$ and $m_*=m$, and \eqref{eq:asy-small-comparison} implies $m\leq P_{\varepsilon\kappa}(\Phi)\leq\E_{\Q_\varepsilon}[\Phi]\to m$. If it is empty, the attained minimum $j_*$ is positive and $P_{\varepsilon\kappa}(\Phi)\leq O-j_*/\varepsilon\to-\infty$. The monotonicity in both limits follows from Proposition~\ref{prop:asy-calculus}\textup{(iii)}.
\end{proof}

\begin{proof}[Proof of Proposition~\ref{prop:rate-OT}]
By \eqref{eq:asy-OT-comparison}, for every $\varepsilon>0$,
$$I_\kappa(\Q_\varepsilon) \leq\varepsilon\bigl(O-P_{\varepsilon\kappa}(\Phi)\bigr)\leq j_O.$$
The proof of Theorem~\ref{thm:asy-endpoints}\textup{(i)} shows that $I_\kappa(\Q_\varepsilon)\to j_O$ as $\varepsilon\to\infty$. Hence $\varepsilon\bigl(O-P_{\varepsilon\kappa}(\Phi)\bigr)\to j_O$ as $\varepsilon\to\infty$, which proves \eqref{eq:rate-OT-first-order}.
\end{proof}

\begin{proof}[Proof of Proposition~\ref{prop:rate-additive}]
Fix $\varepsilon>0$ and $(\varphi,H)$ as in \eqref{eq:rate-hedge}. With $H_{-1}=H_N=0$, each order satisfies $|\Delta H_i^a|\leq C_i^a+C_{i-1}^a$ for $0\leq i\leq N$, including the opening and liquidation orders. The impact cost is thus bounded on $\Omega$, and the trading gain before impact has at most linear growth in the price path. Hence $H$ satisfies \eqref{eq:admissible} at impact $\varepsilon\kappa$.

Fix $\Q\in\Pi(\mu)$. Since $H$ is admissible, its expected net gain $\E_\Q[G_{\varepsilon\kappa}(H)]$ is at most the supremum $I_{\varepsilon\kappa}(\Q)$ in \eqref{eq:penalty}. Subtract $G_{\varepsilon\kappa}(H)$ from the hedge inequality in \eqref{eq:rate-hedge}. The frictionless trading gain cancels by \eqref{eq:gain}, and the static payoffs have total expectation $m$ under $\Q$. Taking expectations, we obtain
\begin{align*}
 \E_\Q[\Phi]-I_{\varepsilon\kappa}(\Q)
 &\leq \E_\Q[\Phi-G_{\varepsilon\kappa}(H)]\\
 &\leq m+\frac{\varepsilon}{2}\E_\Q
       \sum_{i=0}^N\sum_{a=1}^d\kappa_i^a(\Delta H_i^a)^2\\
 &\leq m+\frac{\varepsilon}{2}
       \sum_{i=0}^N\sum_{a=1}^d\kappa_i^a(C_i^a+C_{i-1}^a)^2.
\end{align*}
Taking the supremum over $\Q\in\Pi(\mu)$ proves the upper bound in \eqref{eq:rate-smooth}. For the lower bound, a martingale law attaining $m$ has zero penalty by Theorem~\ref{thm:main}\textup{(ii)}. Its objective value in \eqref{eq:primal} is $m$, hence $P_{\varepsilon\kappa}(\Phi)\geq m$.
\end{proof}

\keeprevisioncolorsfalse%

\appendix
\section{Calibration and numerical implementation}\label{app:fin-calibration}
\subsection{Lognormal forward-start call}\label{app:fin-grid}

For the reference price in Section~\ref{subsec:fin-simulation}, the log returns over the two successive periods are independent and Gaussian. Write $R=X_2/X_1$ and $w=v_2-v_1$. Then $R$ is independent of $X_1$ and $\log R\sim N(-w/2,w)$. Let $F_{\rm N}$ denote the standard normal distribution function. Since the payoff is $X_1(R-1)^+$ and $\E[X_1]=1$,
$$\E[(X_2-X_1)^+] = \int_{\sqrt{w}/2}^{\infty} \bigl(e^{\sqrt{w}\,z-w/2}-1\bigr)\,dF_{\rm N}(z) = 2F_{\rm N}(\sqrt{w}/2)-1.$$
With $w=v_2-v_1 = 1\times 0.175^2-0.5\times 0.195^2 = 0.0116125$, the expected payoff is $429.70$ basis points of the initial spot.

To solve \eqref{eq:fin-wot-price} numerically, we use 256 support points for each marginal. Both dates use the percentile cutoffs $u_j=\sin^2(\pi j/512)$, $j=0,\ldots,256$. At date $T_i$, we place the mass $p_j=u_j-u_{j-1}$ at the average price between the $u_{j-1}$ and $u_j$ quantiles of $X_i$. For the lognormal law in Section~\ref{subsec:fin-simulation}, this support point is
$$x_{ij}= \frac{F_{\rm N}\!\left(F_{\rm N}^{-1}(u_j)-\sqrt{v_i}\right) - F_{\rm N}\!\left(F_{\rm N}^{-1}(u_{j-1})-\sqrt{v_i}\right)} {p_j}, \qquad j=1,\ldots,256,$$
with $F_{\rm N}^{-1}(0)=-\infty$ and $F_{\rm N}^{-1}(1)=+\infty$. These conditional averages preserve $\E[X_i]=1$ at each date. We check convex order by comparing call prices at every strike in the union of the two discrete supports.

For each positive $\eta$, the objective in \eqref{eq:fin-wot-price} evaluated at a feasible coupling bounds the seller's price from below. For the upper estimate, we fix the computed holding and cover the remaining payoff with static positions at every pair of support points. The two estimates differ by less than $2.2\times10^{-7}$ basis points. Table~\ref{tab:fin-bs} reports their midpoint and the expected number of shares traded per claim computed from \eqref{eq:fin-single-volume}. The first and last rows show the MOT and OT upper values for the discrete marginals.

\begin{table}[H]
 \centering\small
 \begin{tabular}{lrr}
  \hline
  $\eta$ & Seller's upper bound (bp) & Expected shares traded per claim \\
  \hline
  $0$ (MOT)      & $537.54$      & -- \\
  $0.0001$       & $537.75$      & $1.037$ \\
  $0.0003$       & $538.16$      & $1.036$ \\
  $0.001$        & $539.59$      & $1.030$ \\
  $0.003$        & $543.60$      & $1.014$ \\
  $0.01$         & $556.76$      & $0.967$ \\
  $0.03$         & $589.14$      & $0.875$ \\
  $0.1$          & $671.93$      & $0.715$ \\
  $0.3$          & $817.49$      & $0.548$ \\
  $1$            & $1{,}038.11$  & $0.309$ \\
  $3$            & $1{,}160.02$  & $0.130$ \\
  $10$           & $1{,}215.02$  & $0.044$ \\
  $30$           & $1{,}233.22$  & $0.016$ \\
  $100$          & $1{,}240.08$  & $0.005$ \\
  $\infty$ (OT)  & $1{,}243.15$  & -- \\
  \hline
 \end{tabular}
 \caption{Forward-start call with $k=1$ and discretized lognormal
 marginals. Prices are in basis points of the initial spot. The share
 count includes opening at $T_1$ and liquidation at $T_2$.}
 \label{tab:fin-bs}
\end{table}

\subsection{SPX calibration and price results}\label{subsec:fin-calibration}

For the two maturities in Section~\ref{subsec:fin-empirical}, we use Cboe quotes with positive bids and asks at least as high as the corresponding bids.\footnote{The snapshot from the \href{https://cdn.cboe.com/api/global/delayed_quotes/options/_SPX.json}{Cboe SPX option chain} is timestamped \texttt{2026-10-09 02:25:46} and records the last trade in the underlying on 8 October.} For the parity fit, we match calls and puts at strikes between $85\%$ and $115\%$ of the spot recorded in the snapshot. We regress the call midpoint minus the put midpoint on their common strike $K$ by least squares. The fitted line has slope $-D_i$ and crosses zero at $F_i$.

For the marginal fit, we use puts with $K<F_i$ and calls with $K\ge F_i$, restricting $K/F_i$ to $[0.70,1.30]$. We require positive displayed sizes on both sides and open interest of at least 100 contracts. The spread must be at most two index points and at most $5\%$ of the midpoint.

For both maturities, we take 181 values of the normalized index from $0.55$ to $1.45$ in steps of $0.005$. This interval covers all normalized strikes used in the fit. We add seven nodes at $0$, $0.2$, $0.35$, $0.45$, $1.6$, $1.8$, and $2$ to allow probability mass farther out in the tails. Let $x_1,\ldots,x_{188}$ denote these nodes and $p_{ij}$ the mass assigned to $x_j$ at $T_i$. We impose
\begin{equation}\label{eq:fin-calibration-constraints}
 \begin{gathered}
  p_{ij}\ge 0,\qquad \sum_j p_{ij}=1,\qquad
  \sum_j x_jp_{ij}=1,\qquad i=1,2,\\
  \sum_j (x_j-x_r)^+(p_{2j}-p_{1j})\ge 0,
  \qquad r=1,\ldots,188.
 \end{gathered}
\end{equation}
The second line imposes the call price ordering in Section~\ref{subsec:fin-empirical} at each node. Between consecutive nodes, the difference between the fitted December and November call prices is affine in the normalized strike. It is nonnegative at both endpoints and hence throughout the interval. For a strike $u\le 0$, both prices equal $1-u$, and for $u\ge 2$, both are zero. The call price characterization of convex order therefore implies $\mu_1\cv\mu_2$.

We multiply each expected normalized payoff by $D_iF_i$ to express its fitted price in index points. For each quote in the sample, we divide the absolute difference between the fitted price and the midpoint by the larger of half the quoted spread and $0.025$ index points. We first minimize the average of these scaled errors. Among fits whose average is within $10^{-8}$ of the minimum, we choose a pair with the smallest sum of second moments. Table~\ref{tab:fin-calibration} reports the results by maturity. Every fitted price lies within the quoted bid and ask interval.

\begin{table}[!htbp]
 \centering\small
 \begin{tabular}{lrrrrr}
 \hline
 Expiry & $F_i$ & $D_i$ & Quotes & Support points & RMSE \\
 \hline
 20 Nov 2026 & $7815.1567$ & $0.9949925$ & 262 & 47 & $0.0531$ \\
 18 Dec 2026 & $7835.0209$ & $0.9916631$ & 296 & 64 & $0.0536$ \\
 \hline
 \end{tabular}
 \caption{SPX forward prices, discount factors, and pricing errors by expiry. Support points have positive fitted mass. RMSE is the root mean squared difference
 between fitted prices and quote midpoints, in index points.}
 \label{tab:fin-calibration}
\end{table}

Table~\ref{tab:fin-empirical-prices} reports the bounds per claim in \eqref{eq:fin-quantity} at $k=1$ for the three order sizes in Figure~\ref{fig:fin-empirical}. We compute the expected number of shares traded under a law attaining the seller's upper bound and multiply the volume per claim by $n$ to obtain the volume for the whole order.

\begin{table}[!htbp]
 \centering\small
 \begin{tabular}{lrrrr}
 \hline
 & \multicolumn{2}{c}{Price per claim (bp)}
 & \multicolumn{2}{c}{Expected shares traded} \\
 $n$ & Lower & Upper & Per claim & Whole order \\
 \hline
 $n\downarrow 0$ (MOT)   & $69.27$ & $208.37$ & --      & --      \\
 $0.1$                   & $69.18$ & $209.73$ & $0.829$ & $0.083$ \\
 $1$                     & $68.48$ & $220.66$ & $0.741$ & $0.741$ \\
 $10$                    & $65.72$ & $281.60$ & $0.480$ & $4.802$ \\
 $n\uparrow\infty$ (OT)  & $59.63$ & $401.17$ & --      & --      \\
 \hline
 \end{tabular}
 \caption{SPX forward-start call with $k=1$ and $\eta_0=0.01$.
 Price bounds are in normalized basis points. Trading volume
 includes opening at $T_1$ and liquidation at $T_2$.}
 \label{tab:fin-empirical-prices}
\end{table}

\subsection{Computations for the basket call}\label{app:basket-implementation}

For the basket call in Section~\ref{subsec:fin-basket}, let $q(x,y)\ge0$ be the mass assigned to the price vectors $x$ at $T_1$ and $y$ at $T_2$. For each of the four coordinates and each of its 16 midpoints, the sum of $q(x,y)$ over the other three coordinates must equal $1/16$. These constraints allow the probability $p(x)=\sum_y q(x,y)$ of observing the pair $x$ at $T_1$ to vary.

Write $d(x)=\sum_y q(x,y)(y-x)$. When $p(x)>0$, the conditional drift of asset $a$ given $X_1=x$ is $d^a(x)/p(x)$. Since impact costs add across assets, the seller's upper bound in \eqref{eq:primal} for $\eta_1,\eta_2>0$ is
\begin{equation}\label{eq:fin-basket-discrete}
 \max_q\left\{
   \sum_{x,y} q(x,y)(y^1+y^2)^+
   -\sum_{x:\,p(x)>0}\sum_{a=1}^2
      \frac{d^a(x)^2}{2\eta_a p(x)}
 \right\}.
\end{equation}
For the OT upper bound we maximize the payoff term alone. For the MOT upper bound, we also require $d(x)=0$ at every $x$.

In the computation we replace the penalty by $\sum_{x,a}t_a(x)$ with $t_a(x)\ge0$ and $d^a(x)^2\le 2\eta_a p(x)t_a(x)$. After solving the resulting second-order cone program, we adjust the masses to match the marginals. The objective in \eqref{eq:fin-basket-discrete} at these masses is a lower estimate of the maximum. For an upper estimate, we fix the holdings returned by the solver and maximize the expected payoff less the net trading gain over the same feasible laws. This is a linear program in $q$. We report the midpoint of these two numerical values. For all positive impact coefficients in Figure~\ref{fig:fin-basket-prices}, they differ by less than $8.1\times10^{-6}$ in payoff units.

Under the computed joint law, the holding in asset $a$ at $T_1$ is $d^a(x)/(\eta_a p(x))$ shares when $X_1=x$ and $p(x)>0$. Counting the opening trade and liquidation as in \eqref{eq:fin-single-volume}, we obtain an expected volume of $(2/\eta_a)\sum_x |d^a(x)|$ shares per claim in that asset. For $(\eta_1,\eta_2)=(0.1,0.01)$, the expected volumes are $0.31$ shares in the first asset and $3.73$ in the second. Interchanging the coefficients reduces these volumes to $0.23$ and $1.07$ shares, respectively.

\end{document}